\documentstyle[12pt]{article}
\begin{document}
\typeout{--- Title page start ---}

\thispagestyle{empty}
\renewcommand{\thefootnote}{\fnsymbol{footnote}}

\begin{tabbing}
\hskip 11.5 cm \= {Imperial/TP/92-93/21}
\\
\> {astro-ph@babbage.sissa.it --}
\\
\>{-- 9303001}
\\
\> October 1993 \\
\> Submitted to PRD \\
\end{tabbing}
\vskip 1cm
\begin{center}
{\Large\bf Inflation and squeezed quantum states} \\
\vskip 1.2cm
{\large\bf Andreas Albrecht, Pedro Ferreira, Michael Joyce
\footnote{Address from Jan. 1st 1993: Physics Dept., Princeton
University, Princeton, N.J. 08544, USA.}
Tomislav Prokopec$^{*}$}
\\
Blackett Laboratory, Imperial College\\
Prince Consort Road, London SW7 2BZ  U.K.
\end{center}
\vskip 1cm
\begin{center}
{\large\bf Abstract}
\end{center}

The inflationary cosmology is analyzed from the point of view of
squeezed quantum states. As noted by Grishchuk and Sidorov,
the amplification of quantum fluctuations into macroscopic perturbations
which occurs during cosmic inflation is a process of quantum
squeezing. We carefully develop the squeezed state formalism and derive the
equations that govern the evolution of a gaussian initial state.
We derive the power spectrum of density perturbations for a simple
inflationary model and discuss its features.
We conclude that the squeezed state formalism provides an
interesting framework within which to study the amplification
process, but,in disagreement with the claims of Grishchuk and Sidorov
, that it does {\em not} provide us with any new physical results.

\vskip 1cm
\typeout{--- Main Text Start ---}

\renewcommand{\thefootnote}{\arabic{footnote}}
\setcounter{footnote}{0}

\section{Introduction}
One of the impressive features of an inflationary cosmology is the
prediction of a set of perturbations on the background
Robertson-Walker
metric. These perturbations are produced {\it via\/} the amplification of
ground state quantum fluctuations during
the inflationary period. This process has been widely studied
and there is broad agreement
regarding both methods and results [\ref{original}]. The
actual perturbations predicted depend on details of the
inflationary period.   A cosmology with a period of simple
exponential inflation and with cold dark matter forms the basis
of the ``standard CDM'' model for
the formation of galaxies and other structure in the universe. This
model has enjoyed great popularity, but it is also coming under increasing
pressure from astronomical observations
[\ref{Smoot}],[\ref{Lubin}],[\ref{Gorski}],[\ref{EfstathiouFrank}].

Recent work by Grishchuk and Sidorov [\ref{scalarperta}],
[\ref{scalarpertb}] has suggested that
important quantum effects have been neglected in the standard
approach. These authors claim that, because of quantum squeezing,
inflation predicts features in the perturbations which have not
been properly taken into account and which could result in striking
observational consequences. In particular, they emphasize the
phenomenon of desqueezing which leads to approximate zeros in the power
spectrum at calculable wavelengths.

We have systematically investigated the inflationary cosmology from
the point of view of quantum squeezing, using Bardeen's gauge invariant
variables [\ref{Bardeen}]. We
have found that indeed each mode of the perturbed field evolves as a
squeezed state during the inflationary period
but that the features discussed by Grishchuk and Sidorov
in [\ref{scalarperta}] and [\ref{scalarpertb}] are well known ones,  which
are essentially classical in nature.  Although we note in section 6 an
isolated error in the literature which may have prompted much of Grishchuk's
criticism, we argue that the error can be (and usually is) avoided without
appealing to the formalism of squeezed quantum states.
We conclude that this
perspective offers nothing more then an alternative
 set of words (and variables)
with which to discuss the inflationary universe. We do however
find the squeezed state formalism well suited to
the problem [\ref{BrandenbergerMukhanovProkopec}] and it may prove
useful in future work.

The structure of the paper is as follows. In section 2 we look at a
simple mechanical system -- the inverted harmonic oscillator -- and show
how it exhibits squeezing behavior at both the classical and
quantum levels. In sections 3 and 4 we use the formalism of gauge
invariant cosmological perturbations as
presented in [\ref{MukhanovFeldmanBrandenberger}], to
construct the Hamiltonian operator. We then set up the
time evolution operator and show  that it can be factorized into a product of a
squeeze operator and a rotation operator which are characterized in
terms of the squeeze factor $R_{\vec{k}}$, squeeze phase
$\Phi_{\vec{k}}$ and the rotation angle $\Theta_{\vec{k}}$.
$R_{\vec{k}}$ gives us a measure of the excitation of the state while
$\Phi_{\vec{k}}$ gives us a measure of how the excitation is shared
between canonical variables. We show how the evolution of the state
can be characterized by a set of coupled first order ordinary
differential equations for $R_{\vec{k}}$, $\Phi_{\vec{k}}$ and
$\Theta_{\vec{k}}$. In section 5  we study
the behavior of this system of ODE's, identifying different regimes
according to the scale of the perturbations: on scales larger then
the Hubble radius the squeeze phase freezes out and the squeeze factor
grows; on scales smaller than the Hubble radius the squeeze parameters
oscillate.

Having gained some insight into what to expect generically
in such models we look at a simple inflationary model with baryonic matter
coupled to photons (without dark matter) such that the evolution of
perturbations can be well approximated by a single collective scalar
field.  We generate some typical power spectra, $|\delta_k|^2$, and see that
they are Harrison-Zeldovich on superhorizon scales ($|\delta_k|^2 \propto
k$ -- no oscillations) and  exhibit standard sound wave
oscillations on subhorizon scales.

In section 6 we discuss the desqueezing effect
emphasized by Grishchuk and Sidorov and argue that it a familiar one
properly taken into account in standard calculations. In section 7
we attempt to clarify the claim that these effects are of
a distinctly quantum mechanical origin. We comment,
using the language of squeezed states, on the classicality of the
harmonic oscillator; we note that, for large squeezing, the
squeezed state satisfies the WKB
criterion for classicality. This is equivalent to
the WKB classicality at late times in an inverted harmonic oscillator
studied by Guth and Pi in Ref. [\ref{GuthPi}].  The point of this
section is to explain that the apparently very quantum mechanical
squeezed state is in fact classical in the sense with which cosmologists
are familiar. That the truly quantum mechanical features of these
states which are probed, for example, in quantum optics might have
cosmological implications is a fascinating claim but one which has
no substance at present. In section 8 we summarize briefly and
conclude.

\section {The Single Inverted Harmonic Oscillator}
\label{tsiho}

The aim of this section is to familiarize the reader with the language
of squeezed states. We apply the squeezed state formalism to a simple
system
 -- the inverted harmonic oscillator. We will
show first how this system exhibits squeezing behavior
at the classical level. We show how this behavior is due to the
presence of one growing and one decaying solution and that essentially
the same behavior carries over to the quantum mechanical system.
\subsection{\it Classical}
The inverted harmonic oscillator (with unit mass and spring constant)
is described by the Hamiltonian:
\begin{eqnarray}
H={p^2 \over 2}-{{q^2}\over 2}.
\label{hiho}
\end{eqnarray}
A convenient choice of variables is
\begin{eqnarray}
b_{\pm} \equiv {1 \over \sqrt{2}}(p \pm q).
\end{eqnarray}
The general solutions are:
\begin{eqnarray}
b_+(t) = b_+(0)e^{t} & b_-(t) = b_-(0)e^{-t}.
\end{eqnarray}
The evolution of the inverted harmonic oscillator is illustrated
in Fig 1,
which shows the trajectories in phase space of a few representative
solutions. The phase space can be labeled equally well by $p$ and $q$
or $b_+$ and $b_-$ (the rotated axes). As time goes on the value of $b_+$
gets exponentially large, while the value of $b_-$ gets exponentially
small. This is because all (but one) of the solutions eventually go
``rolling down the hill''. As this occurs, $p$ and $q$ each grow
exponentially, while their difference exponentially approaches zero.

The trajectories in Fig 1 describe squeezing in the sense that they
get closer together in the $b_-$ direction and further apart in the $b_+$
direction. For example, the circle in Fig 1 evolves into
the squeezed
shape above it after a period of time.  Any probability
distribution in phase space will eventually become squeezed along the $p=q$
axis
as the system evolves.
\subsection{\it Quantum}
Now consider the quantum system described by Eq. \ref{hiho}. Using
the usual $\hat{a}$ and $\hat{a}^{\dagger}$ defined for the
right-side-up harmonic oscillator we find that
\begin{eqnarray}
H &=& {\hat{p}^2 \over 2}-{{\hat{q}^2}\over 2} \\
&=& i{{\hbar } \over 2}(\hat{a}^2e^{2i{\pi \over 4}}-h.c.).
\end{eqnarray}
We have written the Hamiltonian in this way because this is the
form directly comparable with the more general squeeze Hamiltonian
which we will consider.

If the system starts in the vacuum state annihilated by $\hat{a}$
(which is just the gaussian ground state of the right-side-up
oscillator) it evolves into a ``squeezed state'' given by
\begin{eqnarray}
|\Psi(t) \rangle= {\cal S}|0 \rangle=e^{{r \over
2}({\hat{a}}^2e^{-2i\phi}-h.c.)}|0 \rangle.
\label{fg}
\end{eqnarray}
The ``squeeze operator'' ${\cal S}$ is specified by two parameters:
$r$, the ``squeeze factor'', and $\phi$, the ``squeeze phase''. For a general
squeeze operator $r$ and $\phi$ can be complicated functions of time,
but in this simple case they
reduce to $r= t$ and $\phi=-\pi /4$.

We now discuss the squeezed state in connection with the Heisenberg
uncertainty relationship. Using the
relation
\begin{eqnarray}
{\cal S}^{\dagger}a{\cal S}={\hat{a}}\cosh r
-{\hat{a}^{\dagger}}e^{2i\phi}\sinh r,
\end{eqnarray}
it can be easily shown that
\begin{eqnarray}
\hat{p}|\Psi(t) \rangle=
\alpha(r,\phi)\hat{q}|\psi(t)\rangle,
\label{alpha}
\end{eqnarray}
where
\begin{eqnarray}
\alpha(r,\phi)=i{{\cosh r +e^{2i\phi}\sinh r} \over {\cosh r -
e^{2i\phi}\sinh r}}.
\end{eqnarray}
It then follows that
\begin{eqnarray}
\langle \Psi(t)|\hat{p}^2 |\Psi(t) \rangle &=&
|\alpha(r,\phi)|^2 \langle \Psi(t)|\hat{q}^2|\Psi(t)\rangle \\
&=& {{\hbar \over  2 }}(\cosh 2r +\sinh 2r \cos 2 \phi),
\label{psquared}
\end{eqnarray}
and the uncertainty relationship is
\begin{eqnarray}
\left(\langle \Psi(t)|\hat{q}^2|\Psi(t) \rangle \langle
\Psi(t)|\hat{p}^2|\Psi{t}\rangle \right)^{1 \over 2}={\hbar \over 2}(1
+\sin^2 2 \phi \sinh^2 2r)^{1 \over 2}.
\end{eqnarray}
Thus $\Delta q\Delta p \simeq {1 \over 4}\hbar e^{2  t}$ for $ t
\gg 1$. The initial minimum uncertainty gaussian state which ``sits at
the top of the hill'' spreads rapidly in $q$ and $p$.

Consider however
\begin{eqnarray}
\langle \Psi(t)|(\hat{p} \cos \phi - \hat{q} \sin \phi )^2
|\Psi(t)\rangle ={\hbar \over 2}e^{2r}
\end{eqnarray}
and
\begin{eqnarray}
\langle \Psi(t)|(\hat{p} \sin \phi +  \hat{q} \cos \phi )^2
|\Psi(t)\rangle ={\hbar \over 2}e^{-2r}.
\end{eqnarray}
For $\phi = -\pi/4$ these are just $(\Delta b_+)^2$ and $(\Delta b_-)^2$.
Thus in the  $p-q$ plane we say that the
state is squeezed along an axis with slope $\tan \phi$. The
fluctuations normal to this axis are exponentially small. This
behavior mirrors that of phase space trajectories
for the classical
system (see Fig. 1) and likewise corresponds to the existence
of one decaying and one growing solution.

The state can in fact be represented as a phase space density, using the
Wigner function [\ref{Halliwell}],
for which the contours are ellipses with one axis of
length $e^{2r}$ defined by the angle $\phi$ and the other axis of
length $e^{-2r}$ as in Fig. 1. The squeezed states which we will
consider will have a time dependent $\phi$ so they can be pictured as ellipses
rotating in the phase space.

Quantum squeezed states generate considerable interest in various
areas of physics, {\it e.g.\/} nonlinear optics
[\ref{Caves}], [\ref{quantumoptics}], gravity
waves [\ref{Sidorov}], [\ref{GSgravwaves}], gravity wave
detectors, and quantum cosmology
[\ref{Zurek}]. Their striking
feature is that they exhibit dramatically the Heisenberg uncertainty
relation, by allowing one variable to have arbitrarily small
uncertainty. The conjugate variable has a compensating large
uncertainty so the Heisenberg uncertainty relation is obeyed as an equality.
In this
sense squeezed states are very quantum mechanical.  We
will discuss the issue of classical {\it vs.\/} quantum aspects in more detail
in section 7.

\newpage
\section{Formalism For Cosmological Perturbations }
The gauge invariant formalism of cosmological perturbations is well
suited to the study of the evolution of vacuum fluctuations.
As discussed in [\ref{MukhanovFeldmanBrandenberger}], the problem
is reduced to the analysis of the evolution of a scalar
field with a time dependent mass.

If one looks solely at the scalar degrees of freedom of the metric
perturbations
\begin{eqnarray}
 \delta g_{\mu \nu}=a^2(\eta) \left( \begin{array}{clcr} 2 \phi & -B_{|i} \\
-B_{|i} & 2(\psi \gamma_{ij}-E_{|ij}) \end{array} \right) \, ,
\end{eqnarray}
it is possible to combine the functions $\phi,\psi,E,B$ into two gauge
invariant quantities (invariant under
local coordinate transformations)
\begin{eqnarray}
 \Psi=\psi- {\cal H}(B-E')\, , & \quad \Phi= \phi +(1/a)[(B-E')a]'\, ,
\end{eqnarray}
where ${\cal H}= {a'/a}$ is the conformal Hubble parameter,
$a$ denotes the scale factor and $'=d/d \eta$ denotes the
derivative with respect to conformal time.
We can do the same thing with the matter fields; for
example with a scalar field,
$\varphi( \vec{x}, \eta)=\varphi_0 (\eta) +\delta \varphi
(\vec{x},\eta)$, we can build a gauge
invariant quantity
\begin{eqnarray}
\delta \varphi^{(gi)}=\delta \varphi +\varphi_0'(B-E')\, .
\end{eqnarray}
These gauge invariant quantities can be combined into a single scalar field
\begin{eqnarray}
v=a(\delta \varphi_{matt}^{(gi)}+z\Psi)\, ,
\end{eqnarray}
where $\delta \varphi^{(gi)}_{matt}$ denotes a generic
matter field perturbation, $z$ is given by
\begin{eqnarray}
z=({a/{c_s {\cal H}}})[{{2 \over 3}({\cal H}^2-{\cal H}')}]^{1/2},
\label{zdef}
\end{eqnarray}
and $c_s=(\delta p_0 /\delta \epsilon_0)^{1 / 2}$ denotes the speed of sound
(in inflation the correct equations are obtained by setting $c_s \equiv 1$).
The action for the perturbations can then be written as
\begin{eqnarray}
S_{\rm pert}={1 \over 2}
\int d^4x[(v')^2-c_s^2(v_{,i})^2+{z'' \over z}v^2],
\label{eq:actionMFB}
\end{eqnarray}
which is the action for a free scalar
field $v$ with a time dependent mass ($m^2=-z''/z$)
[\ref{MukhanovFeldmanBrandenberger}].
Up to a total derivative term this action is equivalent to the action
\begin{eqnarray}
S^{\prime}_{\rm pert}={1 \over 2} \int d^4x [(v')^2-c_s^2(v_{;i})^2-
2{z' \over z}vv'+({z' \over z})^2v^2]\, ,
\label{eq:actionG}
\end{eqnarray}
which we will find more convenient to work with.
We can now proceed with the standard quantization. Constructing the Hamiltonian
we get
\begin{eqnarray}
 H={1 \over 2}\int d^{3}x[\pi^2+c_s^2{(v_{,i})}^2+2{z' \over z} v \pi]\, .
\label{eq:HamG}
\end{eqnarray}
Promoting the fields to operators and taking the Fourier decomposition so that
\begin{eqnarray}
\hat{v} & = & \int{d^3k \over {(2 \pi)^{3 / 2}}}
 \hat{v}_{\vec{k}} e^{i\vec{k} \cdot \vec{x}} \nonumber \\
\hat{\pi} & = &
\int{d^3k \over {(2 \pi)^{3 / 2}}}\hat{\pi}_{\vec{k}} e^{i\vec{k} \cdot
\vec{x}}\, ,
\end{eqnarray}
we get the two-mode Hamiltonian
\begin{eqnarray}
\hat{{\cal H}}_{\vec{k}}=
\hat{\pi}_{- \vec{k}}\hat{\pi}_{\vec{k}}+
c_s^2k^2\hat{v}_{- \vec{k}}\hat{v}_{\vec{k}}+
{z' \over z}(\hat{\pi}_{- \vec{k}}\hat{v}_{\vec{k}}+
\hat{v}_{- \vec{k}}\hat{\pi}_{\vec{k}})\, .
\end{eqnarray}

We want to work in the Schr\"{o}dinger picture, in which
the operators
$\hat v_{\vec k}$ and $\hat \pi_{\vec k}$ are fixed at an initial
time. We define modes with initial frequency equal to $k$ which,
suitably normalized, give
\begin{eqnarray}
\hat{v}_{\vec{k}} & =
& {1 \over \sqrt{2k}}({a_{\vec{k}}+a^{\dagger}_{- \vec{k}}}) \nonumber \\
\hat{\pi}_{\vec{k}} & = & - i \sqrt{k \over
2}({a_{\vec{k}}-a^{\dagger}_{- \vec{k}}})\,.
\label{eq:vPiSch}
\end{eqnarray}
The two-mode Hamiltonian operator can be written in the simple form
\begin{equation}
\hat{{\cal H}}_{\vec{k}}=
\hat{{\cal H}}_{\vec{k}}^{(0)}+\hat{{\cal H}}_{\vec{k}}^{(I)}=
\Omega_{\vec{k}}(a^{\dagger}_{\vec{k}}a_{\vec{k}}+
a^{\dagger}_{- \vec{k}}a_{- \vec{k}}+1) +
i\lambda_{\vec{k}} (e^{-2i\varphi_{\vec{k}}}
a_{\vec{k}}a_{- \vec{k}}-{\rm h.c.\/}) \, ,\label{eq:Ham}
\end{equation}
where
\begin{eqnarray}
\Omega_{\vec{k}} & = &{k \over 2}(1+c_s^2) \nonumber \\ \lambda_{\vec{k}} & = &
\left[{\left({k \over 2}(1-c_s^2)\right)^2+\left({{z'}
\over z}\right)^2}\right]^{1 \over 2}
 \nonumber \\
\varphi_{\vec{k}} & =
& -{\pi \over 2}+{1 \over 2}\arctan \left({{{{k z} \over 2z'}(1-c_s^2)}}
\right)\, .\label{eq:OmegaLambdaPhi}
\end{eqnarray}

Eqs. (\ref{eq:Ham})  and (\ref{eq:OmegaLambdaPhi})  describe the
generic momentum
conserving quadratic
Hamiltonian for a scalar field.
It has a free evolution piece, ${\cal H}^{\:\,\, (0)}_{\vec{k}}$
with a time dependent frequency $\Omega_{\vec{k}}$,
and a squeezing piece, ${\cal H}_{\vec{k}}^{\;\,\, (I)}$, with a coupling
strength $\lambda_{\vec{k}}(t)$.  The evolution
operator produced by this Hamiltonian
can be factorized in the following way
\begin{eqnarray}
{\cal U}_{{\cal H}_{\vec{k}}}( \eta,\eta_0)=
{\cal S}[R_{\vec{k}},\Phi_{\vec{k}}]
{\cal R}[\Theta_{\vec{k}}]\, ,
\label{eq:UequalSR}
\end{eqnarray}
where ${\cal R}$ is the two-mode rotation operator defined as
\begin{eqnarray}
{\cal R}[\Theta_{\vec{k}}]=
\exp[{-i\Theta_{\vec{k}}({a^{\dagger}_{\vec{k}}a_{\vec{k}}+
a^{\dagger}_{- \vec{k}}a_{- \vec{k}}}+1)}]
\end{eqnarray}
and ${\cal{S}}$ is the two-mode squeeze operator defined as
\begin{eqnarray}
{\cal S}[R_{\vec{k}},\Phi_{\vec{k}}]=\exp \left[{{R_{\vec{k}} \over 2}
(e^{-2i\Phi_{\vec{k}}}a_{- \vec{k}}a_{\vec k}-{\rm h.c.\/})}\right]
\, .\label{eq:SqueezeOp}
\end{eqnarray}
This simple decomposition of the evolution operator is a general property of
momentum preserving quadratic Hamiltonians [\ref{CavesSchumaker}].
The rotation operator alone gives ordinary  oscillations (points in
the phase space of a classical harmonic oscillator rotate about the
origin).  The squeeze operator alone produces squeezing as discussed
in Sect. \ref{tsiho}.
The complete solution to the problem we are considering reduces to
finding $R_{\vec{k}}$, $\Phi_{\vec k}$ and $\Theta_{\vec k}$ as
functions of time. (Note that $\Phi_{\vec k}$ is {\em not} the Bardeen
variable which we shall write  $\Phi^B$!)

\section{Evolution Equations}

In this section we address the generation of cosmological perturbations by
studying the evolution of the initial vacuum state with the
Hamiltonian  discussed in the previous section.

\subsection{\it The Squeezed Vacuum State}

To begin with we have to define the initial conditions of our quantum field
theory.  We assume that all the modes of interest ({\it i.e.\/} the
modes on subhorizon scales today)
are well within the horizon at the initial time. In this case
we have $k|\eta|\gg 1$, which (with $c_s=1$) implies
$\Omega_{\vec k}\simeq k\gg \lambda_{\vec k}\simeq (1/|\eta|)$, and
Eq.~(\ref{eq:Ham}) reduces to the free Hamiltonian
$\hat{\cal H}^{\;\,\, (0)}_{\vec k}$. We then choose for the
initial state the  ground state of the free Hamiltonian, {\it i.e.\/}
the Poincar\`{e} invariant vacuum state, which is defined by
$$
a_{\vec{k}}|0 \rangle _{\rm in}=0\, , \qquad \forall {\vec k}\, .
$$
The action of the rotation operator $\cal R$ produces an irrelevant
phase
$$
{\cal R}[\Theta_{\vec{k}}]|0 \rangle_{\rm in}=
{\rm e}^{i\Theta_{\vec k}}|0 \rangle_{\rm in} \, ,
$$
but, when acted upon by ${\cal S}[R_{\vec{k}},\Phi_{\vec{k}}]$, the
vacuum state transforms  into a two-mode squeezed state [\ref{CavesSchumaker}]
\begin{eqnarray}
|SS_{\vec{k}} \rangle={\cal S}[R_{\vec{k}},\Phi_{\vec{k}}]|0\rangle_{\rm in}=
\sum_{n=0}^{\infty}{1 \over \cosh R_{\vec{k}}}(-e^{2i \Phi_{\vec{k}}}
\tanh R_{\vec{k}})^n|n,{\vec k};n,-{\vec k} \rangle\, ,
\end{eqnarray}
where
\begin{eqnarray}
|n,{\vec k};n,-{\vec k} \rangle=\sum_{n=0}^{\infty}{1\over n!}
(a^{\dagger}_{\vec k}a^{\dagger}_{-\vec k})^n|0\rangle_{in}
\end{eqnarray}
is the two-mode occupation number state.
This part of the evolution operator is responsible for the amplification of the
initial vacuum fluctuations; momentum conserving pairs of quanta are
created. The
squeeze factor is related to the mean number of quanta, $n_{\vec{k}}$,
in the squeezed vacuum state through the relation
$$
  n_{\vec{k}}= \langle SS|\hat{N}_{\vec{k}}|SS\rangle=\sinh^2R_{\vec{k}}\, .
$$

\subsection{\it Evolution Equations}
The problem is to determine the functions $R_{\vec{k}}$, $\Phi_{\vec k}$
and $\Theta_{\vec k}$. The time evolution operator is given by the
time ordered exponential
\begin{eqnarray}
{\cal U}(\eta,\eta_0) & = &
{\cal T}\exp\bigl(-i\int_{\eta_o}^{\eta}d\eta'
{\cal H}_{\vec{k}}(\eta')\bigr)\, \nonumber \\
& = & {\cal T}\exp\Biggl[\int_{\eta_o}^{\eta}d\eta'{\lambda}_{\vec{k}}(\eta')
\Bigl({\rm e}^{-2i\varphi_{\vec k}+2i\int_{\eta^{\prime}}^\eta
\Omega_{\vec k}(\eta'')d\eta''} a_{\vec k}a_{-\vec k}-{\rm h.c.}
\Bigr)\Biggr] \nonumber \\
& &\exp\biggl[-i\int_{\eta_0}^\eta d\eta'\Omega_{\vec k}(\eta')
\bigl( a^{\dagger}_{\vec{k}}a_{\vec{k}}+a^{\dagger}_{- \vec{k}}a_{- \vec{k}}+1
\bigr)\biggr]\, .\qquad\qquad
\label{eq:TimeOrder}
\end{eqnarray}
We divide the  evolution into infinitesimal time intervals
\footnote{
Note that in contrast to the evolution operator which describes parametric
amplification in [\ref{CavesSchumaker}], the time ordering problem is
non-trivial.} $\epsilon$.
The composite property of the evolution operator implies
\begin{eqnarray}
{\cal U}(\eta+\epsilon,\eta_o)=
{\cal U}(\eta+\epsilon,\eta){\cal U}(\eta,\eta_0)\, .
\end{eqnarray}
We can recast this in terms of the squeeze operator ${\cal S}$ and the rotation
operator ${\cal R}$ in the following form
\begin{eqnarray}
{\cal S}[R_{\vec{k}},\Phi_{\vec{k}}]{\,}{\cal R}[\Theta_{\vec{k}}]=
{\cal S}[\delta R_{\vec{k}},\delta \phi_{\vec{k}}]{\,}{\cal R}
[\delta \theta_{\vec{k}}]{\,}{\cal S}[R^0_{\vec{k}},\phi^0_{\vec{k}}]{\,}
{\cal R}[\theta^0_{\vec{k}}]\, .\label{eq:SqRot}
\end{eqnarray}
Taking account of Eq.~(\ref{eq:TimeOrder}) we infer that for small $\epsilon$:
$\delta R_{\vec{k}}\simeq\lambda_{\vec{k}}(\eta)\epsilon$,
$\delta \theta_{\vec{k}}\simeq\Omega_{\vec{k}}(\eta)\epsilon$
and $\delta \phi_{\vec{k}}\simeq\varphi_{\vec{k}}$.
Using the computation properties of the squeeze and
rotation operators, the
right hand side of (\ref{eq:SqRot}) can be written as
\begin{eqnarray}
RHS={\cal S}[\delta R_{\vec{k}},\delta \phi_{\vec{k}}]{\,}
{\cal S}[R^0_{\vec{k}},\phi^0_{\vec{k}}-\delta \theta_{\vec{k}}]
{\,}{\cal R}[\theta^0_{\vec{k}} +\delta \theta_{\vec{k}}]\, .
\end{eqnarray}
In order to express the product of the two squeeze operators in terms of a
single squeeze operator we use the standard composition property, as given in
[\ref{CavesSchumaker}]
$$
{\cal S}[\delta R_{\vec{k}},\delta \phi_{\vec{k}}]{\,}{\cal S}
[R^0_{\vec{k}},\phi^0_{\vec{k}}-\delta \theta_{\vec{k}}]
={\cal S}[R_{\vec{k}},\Phi_{\vec{k}}]{\,}{\cal R}[{\bar{\theta}_{\vec{k}}}]\, ,
$$
where
\begin{eqnarray}
e^{i\bar{\theta}_{\vec{k}}}\cosh R_{\vec{k}} & =
& \cosh R^0_{\vec{k}} \cosh \delta R_{\vec{k}}
\nonumber \\
& & +e^{2i(\phi^0_{\vec{k}}-\delta \phi_{\vec{k}} -\delta \theta
_{\vec{k}})}\sinh R^0_{\vec{k}} \sinh \delta R_{\vec{k}} \nonumber \\
e^{i(2(\Phi_{\vec{k}}-\phi^0_{\vec{k}}+\delta \theta_{\vec{k}})+
\bar{\theta}_{\vec{k}})}\sinh R_{\vec{k}} & = &
\sinh R^0_{\vec{k}} \cosh \delta R_{\vec{k}} \nonumber \\
& & +e^{-2i(\phi^0_{\vec{k}}-\delta \phi_{\vec{k}}-\delta \theta_{\vec{k}})}
\sinh \delta
R_{\vec{k}} \cosh R^0_{\vec{k}}\, .
\end{eqnarray}
For sufficiently small $\epsilon$ we can expand the LHS in
$\delta R_{\vec{k}}$ and $\delta
\theta_{\vec{k}}$ to obtain the recursion relations
\begin{eqnarray}
R_{\vec{k}}(\eta+\epsilon) & = & R_{\vec{k}}(\eta)+\lambda_{\vec{k}}
(\eta)\epsilon \cos 2(\varphi_{\vec{k}}(\eta)
-\Phi_{\vec{k}}(\eta)) \nonumber \\
\Phi_{\vec{k}}(\eta+\epsilon) & =
& \Phi_{\vec{k}}(\eta)-\Omega_{\vec{k}}(\eta)\epsilon+
\lambda_{\vec{k}}(\eta){\epsilon \over 2}
({{\tanh R_{\vec{k}}(\eta) +
\coth R_{\vec{k}}(\eta) }})\sin 2(\varphi_{\vec{k}}(\eta)
-\Phi_{\vec{k}}(\eta)) \nonumber \\
\Theta_{\vec{k}}(\eta+\epsilon) & =
& \Theta_{\vec{k}}(\eta)+\Omega_{\vec{k}}(\eta)\epsilon-\lambda_{\vec{k}}(\eta)
\epsilon {{\tanh R_{\vec{k}}(\eta)}}\sin 2(\varphi_{\vec{k}}(\eta)-
\Phi_{\vec{k}}(\eta))\, ,\label{eq:recursion}
\end{eqnarray}
where $\Theta_{\vec{k}}=\theta^0_{\vec{k}}+\delta
\theta_{\vec{k}} +\bar{\theta}$. The differential form of these
equations is
\begin{eqnarray}
R_{\vec k}' & = & \lambda_{\vec{k}}
\cos2(\varphi_{\vec k}-\Phi_{\vec{k}}) \nonumber \\
\Phi_{\vec {k}}' & = & -\Omega_{\vec {k}}+{\lambda_{\vec {k}} \over 2}
({\tanh R_{\vec {k}}+\coth R_{\vec {k}}})
\sin2(\varphi_{\vec {k}}-\Phi_{\vec {k}}) \nonumber \\
\Theta_{\vec {k}}' & = & \Omega_{\vec {k}}-{\lambda_{\vec {k}} }
{{\tanh R_{\vec {k}}}}\sin2(\varphi_{\vec {k}}-\Phi_{\vec {k}})\, .
\label{eq:eomRPhiTheta}
\end{eqnarray}
These are the equations of motion of our system.
The analogous equations for gravitational waves have been derived in
Ref.~[\ref{Grishchuktwo}]. These can be obtained from
(\ref{eq:eomRPhiTheta}) by specifying $\lambda_{\vec k}=a$,
$\Omega_{\vec k}=k$ and $\phi_{\vec k}=-\pi/4$.

An alternative derivation of these equations is given in Appendix A,
where we use the fact that the mode functions in the Heisenberg picture
can be expressed in terms of the Schr\"{o}dinger picture variables
$R_{\vec k}$, $\Phi_{\vec k}$ and $\Theta_{\vec k}$. We then
show that the Hamilton
equations for the mode functions reduce to those in Eq.
(\ref{eq:eomRPhiTheta}).

\section {Application to a Simple Inflationary model}

Our aim is to study the growth of cosmological perturbations in the
squeeze state formalism for a simple inflationary model.
This section is mostly concerned with studying the solutions to
Eq.~(\ref{eq:eomRPhiTheta}).
In general, when $\lambda_{\vec{k}}$,
$\Omega_{\vec{k}}$ and $\varphi_{\vec{k}}$ are some complicated
functions of time, it is not possible
to solve Eq.~(\ref{eq:eomRPhiTheta}) analytically.
However, before we proceed to a
discussion of the numerical solution, we
can get some insight into the dynamics of the system using
analytical techniques.

\subsection{\it Analytic Approach}

We assume that $\lambda_{\vec{k}}$, $\Omega_{\vec{k}}$ and
$\varphi_{\vec{k}}$ are slowly varying
functions of time, {\it i.e.\/} for $k \eta <1$ we have
$ \Delta \Omega_{\vec{k}} / \Omega_{\vec{k}}$, $\Delta
\lambda_{\vec{k}} / \lambda_{\vec{k}}$,
$\Delta \varphi_{\vec{k}} / \varphi_{\vec{k}}\ll 1$.

In the {\it strong coupling\/} or {\em squeeze dominated}
regime ($\lambda_{\vec{k}}>\Omega_{\vec{k}}$),
the squeeze angle $\Phi_{\vec{k}}$ and the
rotation angle $\Theta_{\vec{k}}$ approach a stable fixed point
(freeze out). The squeeze factor grows monotonically
with time, which reflects the fact that in the course of evolution the
growing mode becomes more and more dominant over the decaying mode.
In the {\it weak coupling\/} regime
($\lambda_{\vec{k}}<\Omega_{\vec{k}}$), the solution is
oscillatory, with the squeeze factor remaining essentially constant
and the dominant features are the oscillations of the squeezed state,
which are revealed physically as the pressure oscillations in the
hydrodynamic fluid.
\vskip 0.1in
\par
{\it (a) Strong coupling : Freeze out}
\vskip 0.1in
For $\lambda_{\vec{k}}>\Omega_{\vec{k}}$ there is a {\it fixed point\/}
(in $\Phi_{\vec{k}}$ and $\Theta_{\vec{k}}$)
to the equations of motion
\begin{eqnarray}
\begin{array}{clcr}
\Phi_{\vec{k}}^{*'} = \Theta_{\vec{k}}^{*'}=0 \, ,&
\quad R_{\vec{k}}'=\lambda_{\vec{k}}\cos
2(\varphi_{\vec{k}}-\Phi_{\vec{k}}^{*})
\label{eq:fixedpoint}
\end{array}
\end{eqnarray}
with
\begin{eqnarray}
\begin{array} {clcr}
\sin 2(\varphi_{\vec{k}}-\Phi_{\vec{k}}^{*})={{2 \Omega_{\vec{k}}}
\over { \lambda_{\vec{k}}}}{(\tanh R_{\vec{k}}
+\coth R_{\vec{k}})^{-1}}
\stackrel{|R_{\vec k}|\gg 1}{\longrightarrow}
{\rm sign} (R_{\vec{k}}) ({\Omega_{\vec{k}}
\over \lambda_{\vec{k}}})\, .
\end{array}
\end{eqnarray}
Using this condition, we can now integrate Eq.~(\ref{eq:fixedpoint})
for the squeeze factor to obtain
\begin{eqnarray}
R_{\vec{k}} \simeq \int
({\lambda_{\vec{k}}^2-\Omega_{\vec{k}}^2})^{1/2}d\eta\, ,
\label{eq:Rapprox}
\end{eqnarray}
so that $R_{\vec{k}}$ grows monotonically.
Most of the squeezing occurs in the strong coupling regime.
\par
\vskip 0.1in
{\it (b) Weak coupling : Oscillations}
\vskip 0.1in
For $\lambda_{\vec{k}} \ll \Omega_{\vec{k}}$
and taking $\lambda_{\vec{k}}$, $\Omega_{\vec{k}}$ and
$\varphi_{\vec{k}}$
constant, we get the solution
$$
\tan(\Phi_{\vec{k}}-\varphi_{\vec{k}})=
\cos \alpha_{\vec k} \tan [-\Omega_{\vec{k}}(\eta-\eta_0)+\alpha_{\vec k}]
-\tan \alpha_{\vec k}
$$
in which $\sin \alpha_{\vec k}=\lambda_{\vec{k}} / \Omega_{\vec{k}}$.
In the case where $\alpha_{\vec k} \ll 1$, this solution reduces to the form
\begin{eqnarray}
\Phi_{\vec{k}} & =
& \varphi_{\vec{k}}-\Omega_{\vec{k}}(\eta-\eta_0) \nonumber \\
R_{\vec{k}} & = & R^0_{\vec{k}}+{\lambda_{\vec{k}}
\over 2 \Omega_{\vec{k}}}\sin 2 \Omega_{\vec{k}}(\eta
-\eta_{0})=R^0_{\vec{k}}+{\lambda_{\vec{k}}
\over {2 \Omega_{\vec{k}}}}\sin 2 (\varphi_{\vec{k}}-
\Phi_{\vec{k}})\, .
\label{eq:RPhioscillatory}
\end{eqnarray}
We can consider this oscillatory solution as a reasonable
approximation  for modes well within the
horizon in both the inflationary era, when
$\lambda_{\vec{k}}/\Omega_{\vec{k}} \simeq 1/k|\eta| \ll 1$,
 and the radiation dominated era, when
$\lambda_{\vec{k}}/\Omega_{\vec{k}} \simeq 1/2$. For
these modes $R_{\vec{k}}$ is constant on average, {\it i.e.\/} there
is no net squeezing and perturbations do not grow.

For modes that cross the horizon during the matter era,
where $\Omega_{\vec{k}} \simeq\lambda_{\vec{k}}$,
we cannot apply this simple analysis.
\vskip 0.1in
{\it (c) An exact solution: The Bunch-Davies vacuum}
\vskip 0.1in
In the exponentially expanding de Sitter stage, when $\Omega_{\vec k}=k$,
$\lambda_{\vec
k}=1/|\eta|$ and $\varphi_{\vec k}=-\pi/2$ there is an exact solution
to the equations of motion (\ref{eq:eomRPhiTheta})
\begin{eqnarray}
R_{\vec k} & = & \sinh^{-1} {1\over{2k\eta}} \nonumber \\
\Phi_{\vec {k}} & = & -{\pi\over 4}-\arctan{1\over{2k\eta}} \nonumber \\
\Theta_{\vec {k}} & = & k\eta+\arctan{1\over{2k\eta}}\, .
\label{eq:BunchDavies}
\end{eqnarray}
This solution corresponds to the Bunch-Davies vacuum [\ref{BirrellDavies}]
, which is an
attractor. If the initial
state (for the modes within the horizon) is not already highly squeezed, one
finds that as the  modes get driven to superhorizon scales they evolve
toward the
Bunch-Davies vacuum. In the language of squeezed state parameters
this corresponds to the freeze-out of $\Phi_{\vec {k}}$ and
$\Theta_{\vec {k}}$; we see this behavior in the limit $k|\eta|\ll 1$
of Eq.~(\ref{eq:BunchDavies}).

\subsection{\it Squeezing in a Simple Inflationary Model}

Having established that most of growth occurs on superhorizon
scales, we now use a simple model to
estimate the amount of squeezing in the perturbation field.
We have found that all of the relevant squeezing
occurs on superhorizon scales, {\it i.e.\/} when $k|\eta|<1$. In
Fig.~2 this corresponds to the interval
$[\eta_{1x},\eta_{2x}]$, where $\eta_{1x}=-1/k$ and
$\eta_{2x} \simeq 1/k$ ($2/k$) in the radiation (matter) era.

The relevant squeezing occurs for the couplings for which
$\lambda_{\vec{k}} \gg \Omega_{\vec{k}}$ when
$\lambda_{\vec{k}} \simeq z'/z$. We can then
integrate Eq.~(\ref{eq:Rapprox}) to obtain
\begin{eqnarray}
R_{\vec{k}}=\int_{\eta_{1x}}^{\eta_{2x}}d\ln z\, .
\label{eq:Rdlnz}
\end{eqnarray}
During the inflationary era (superscript $i$), $z$ can be approximated by
$z(\eta)\simeq ({2/3})^{1/2}a/l_P$
(where  $l_P=(8 \pi G /3)^{1/2}$ is
the Planck length).
The amount of squeezing is given by
$\Delta R^i_{\;\,\vec{k}}\simeq \ln (a_R/a_{1x}) \simeq \ln{(1 / |\eta_R|k)}$.
In the hydrodynamical era, if a mode crosses
 the horizon during the radiation era, then $z(\eta)\simeq
{2}^{1/2}a$ and $\Delta R^{rad}_{\vec{k}}
\simeq \ln (a_{2x} / a_R) \simeq \ln({1 / |\eta_R|k})$.
If the mode crosses the horizon in the matter era
for which $z(\eta) \simeq a^{3/2}{2}^{1/2}$ we have
$\Delta R^{rad}_{\vec{k}}\simeq  \ln({\eta_{\rm eq} / \eta_{R}})$, and
$\Delta R^{matt}_{\vec{k}}\simeq (3/2) \ln(a_{X2} /a_{\rm eq})
\simeq 3 \ln({1 / k\eta_{\rm eq}}).$
This last term is a poor approximation in the matter era; in fact
$\lambda_{\vec{k}}\simeq \Omega_{\vec{k}}$
and the squeezing angle $\Phi_{\vec{k}}$ is not completely frozen
resulting in a slower growth of $R_{\vec{k}}$.
In the case of gravitational waves  (\ref{eq:Rdlnz}) gives
$\Delta R_{\vec{k}} \simeq \ln({a_{2x} / a_{1x}})$,
which is in agreement with the result first obtained by
Grishchuk and Sidorov in [\ref{GSgravwaves}].

\subsection{\it Numerical Analysis}

We shall now study numerically the evolution of perturbations in our
simple model. It is important to point out that during the
hydrodynamic era we are looking only
at the collective field of baryonic matter and
radiation (we are ignoring cold or dark matter, or any other field
which cannot be accurately described by a single collective scalar field).
In addition we ignore  decoupling of matter and
radiation. We do not expect to get results which agree completely
with the highly refined
calculations which  already exist in the literature [\ref{EfstathiouReview}].
However we do expect
approximate agreement if we look solely at the baryonic and
radiation sector of these simulations; in particular in the radiation era
and on superhorizon scales.

The evolution is given by the recursion relations (\ref{eq:recursion})
and we shall assume the following time dependence for the scale factor
\begin{eqnarray}
a_i  = & -{1 \over {H \eta}}\, ,\qquad\qquad\qquad
(-\infty<\eta<-\eta_{R})
&\mbox{inflationary era}\, , \nonumber \\
a  = & {1 \over 4}  \left({{\eta+\theta} \over \eta_{*}} \right)^2
+\left({{\eta+\theta} \over \eta_{*}} \right)\, ,
\quad  (\eta_{R}<\eta<\infty)
&\mbox{hydrodynamical era}\, , \label{adef}
\end{eqnarray}
where $H={\cal{H}}/a=a'/a^2$ is the Hubble constant during inflation, $\theta$
and $\eta_{*}$ are chosen such that
$a(\eta_R)=a_i(\eta_R)$ and $a'(\eta_R)=a_i'(\eta_R)$, where conformal time
$\eta_R$ denotes the end of inflation
(we assume instantaneous reheat).

We normalize $a$ such that
$a_{\rm eq}=1$ and we set $\eta_{*}=1$. As in
[\ref{MukhanovFeldmanBrandenberger}] we assume
$z={{a \varphi_{0}' }/{\cal H}}$ in the inflationary era.
During the early radiation era,
when most of the matter
particles are relativistic, we take $c_s^2=1/3$. We assume
that there is a time,
$\eta=\eta_{\rm rel}$, when matter particles become nonrelativistic.
For $\eta>\eta_{\rm rel}$ we
have $c_s^2=(\delta p_0/ \delta \epsilon_0)_S=1/3(1+3a/4)$.
(We checked that the choice of $\eta_{\rm rel}$ does not influence
squeezing of the state.) The wave function is continuous
at $\eta=\eta_R$, which means that the functions $R_{\vec{k}}$,
$\Phi_{\vec{k}}$ and $\Theta_{\vec{k}}$ are continuous at $\eta_{R}$.
The overall amplitude of the perturbations in the hydrodynamical era
will be dependent on the amount of squeezing in the inflationary era,
which in turn depends on
reheat temperature specified by $\eta_R$.

\vskip 0.1in
\par
{\it Evolution of the squeeze parameters}
\vskip 0.1in
Fig. 3 is a plot of the evolution of the squeeze factor
$R_{\vec{k}}$ as a
function of the scale factor $a$. Most of the growth in the squeeze
factor occurs on the superhorizon scales between the marks {\it 1x\/}
and {\it 2x\/}. When $k \eta \ll 1$, the analytic result discussed
subsequent to
Eq.~(\ref{eq:Rdlnz}) is an excellent approximation:
$R_{\vec{k}} \simeq \ln{{a} / {a_{1x}}}$.

What about subhorizon scales ($k|\eta | >1$)? In inflation
(when $\lambda_{\vec{k}} \ll \Omega_{\vec{k}}$)
$R_{\vec{k}} \simeq 0$, while in the radiation era
$R_{\vec{k}}$ oscillates (see Eq.~(\ref{eq:RPhioscillatory})). For
the modes which enter the horizon in the matter era
(case $k\eta_{\rm eq}=0.1$ on Fig.~2),
the squeeze factor $R_{\vec{k}}$
continues growing as $\Delta R_{\vec{k}} \simeq C_{\vec{k}} \ln{a}$, where
 $C_{\vec{k}} \simeq 1.3$ for
$k \eta_{\rm eq}=0.1$ and $C_{\vec{k}} \rightarrow 1.5$ for
$k \eta_{\rm eq} \ll 1$.
This means that, as a consequence of large coupling in the matter era
($\Omega_{\vec k}<\lambda_{\vec k}$), the squeeze angle remains frozen
($\Phi_{\vec k}\simeq \mbox{const.}$) and the squeezing
continues.
Physically, this is related to the classical process of
gravitational collapse.

There is a critical wave vector,
$k_{\rm crit}\eta\simeq 2$  (which corresponds to the scale
$\lambda_{\rm phys}\simeq
\pi\lambda_{\rm eq}\simeq 40\;(\Omega_0 h^2)^{-1}\mbox{\rm Mpc}$;
$\lambda_{\rm eq}\simeq 13\;(\Omega_0 h^2)^{-1}\mbox{\rm Mpc}$ is
scale today corresponding to the horizon size at equal matter and radiation,
$\Omega_0$ is the fraction of the critical density today and $h$ is
the present Hubble parameter in units of $100\mbox{km/s/Mpc}$).
For $k > k_{\rm crit}$ the state  oscillates
and $R_{\vec{k}}$ doesn't grow, and for
$k < k_{\rm crit}$ the state is frozen  and
$R_{\vec{k}}$ grows.

Fig. 4 shows the evolution of the squeeze factor with respect to
the squeeze phase. On subhorizon scales in inflation $\Phi_{\vec k}$ grows,
while $R_{\vec{k}} \simeq 0$. At the horizon crossing,
the squeeze angle freezes out:
$\Phi^*_{\vec{k}}=n\pi$ ($n\in {\rm Z\!\!\!Z}$) becomes an attractor and,
as we can read of from
the figure: $\Phi^*_{\vec{k}}=0$ ($-2 \pi$) if $k=1$ ($10$) which is in
agreement with Eq.~(\ref{eq:fixedpoint}). For the subcritical case
$k\eta_{\rm eq}=1$, after the mode crosses the horizon at $2x$, the angle
$\Phi\simeq\Phi^*$ remains frozen and $R_{\vec{k}}$ continues growing.
On the other hand, for
$k\eta_{\rm eq}=10 > k_{\rm crit}\eta_{\rm eq}$, after time $\eta_{2x}$ the
mode starts
oscillating with $\Delta R_{\vec{k}} \simeq 1$.
The amplitude of oscillations
in $R_{\vec{k}}$ is slightly bigger than predicted by the simple
formula (\ref{eq:RPhioscillatory}); the reason
being that the condition for validity of Eq.~(\ref{eq:RPhioscillatory})
($\lambda_{\vec{k}} /\Omega_{\vec{k}} \ll1$) is not strictly satisfied
(here we have $\lambda_{\vec{k}} / \Omega_{\vec{k}} \rightarrow 1/2$).
During one oscillation $\Delta\Phi_{\vec{k}}=\pi$ and $R_{\vec{k}}$
remains constant on average. Looking back at Eqs.~(\ref{eq:SqueezeOp})
and (\ref{eq:RPhioscillatory})
we observe that if $\Phi_{\vec{k}}$ grows (oscillations), the squeeze
operator produces and destroys, on average, equal number of particle
pairs, {\it i.e.\/} there is no net squeezing.

\vskip 0.1in
\par
{\it Evolution of physical quantities}
\vskip 0.1in
We are interested in looking at physical quantities in the
hydrodynamical era, typically the
Bardeen   variable $\Phi^B$ (which corresponds to the Newtonian
potential inside the horizon)  and
the energy density perturbations $\delta\epsilon/\epsilon$.
In the standard notation
\begin{eqnarray}
{{\delta \epsilon} \over \epsilon}=
\int{d^3k \over {(2 \pi)^{3 \over2}}} \delta_{\vec{k}} e^{i\vec{k} \cdot
\vec{x}}
\label{eq:deltaepsilon}
\end{eqnarray}
and
\begin{eqnarray}
|\delta_{\vec{k}}|^2 & =
& k\langle {\Phi^{B}_{-\vec{k}}\Phi^B_{\vec{k}}}\rangle
\label{eq:PowerandBardeen} \\
\Phi^B_{\vec{k}} & = & -\sqrt{3 \over 2}l_{P}  {{{\cal H}^2 -{\cal H}'}
 \over {{\cal H}_{\vec{k}}c_s^2}} {1 \over z} {\pi}_{\vec{k}}\, ,
\end{eqnarray}
where $l_P=(8 \pi G /3)^{-{1 \over 2}}$ is the Planck length.
\par

To make contact with the existing work on power spectra from
inflation, we plot in Fig.~5 the growth of the power spectrum
$|\delta_{\vec{k}}|^2$ defined in Eqs.~(\ref{eq:deltaepsilon}) and
(\ref{eq:PowerandBardeen}) against the scale factor for the modes:
$k\eta_{\rm eq}=0.1$ and $k\eta_{\rm eq}=3$. On superhorizon scales,
during the radiation era
($\ln a <0$), the power grows as $|\delta_{\vec{k}}|^2 \propto
a^4\propto \eta^4$,
which agrees with the
estimates based on Eq.~(\ref{eq:Rdlnz}). In the matter era, for the modes
$k\eta_{\rm eq}<2$, the power grows as
$|\delta_{\vec{k}}|^2 \propto a^2\propto\eta^4$,
while for $k\eta_{\rm eq}>2$, the state start oscillating
and the growth becomes very slow.

Fig. 6 shows the power spectrum at two different time
slices: $\eta=0.1\eta_{\rm eq}$ and $\eta=0.5\eta_{\rm eq}$.
The spectrum is scale invariant, $|\delta_{\vec{k}}|^2
\sim k$, on superhorizon scales and
$|\delta_{\vec{k}}|^2 \sim k^{-1}$ on subhorizon scales.
The turning point, caused by the oscillations of the squeezed state
(see Fig. 5), is at $k\eta_{\rm eq} \simeq 4$ for both time slices.
The first dip in the power spectrum is at
$k\eta_{\rm eq} \simeq 8-9$, which corresponds to the wavelength
$\lambda \simeq (0.8-0.9)\lambda_{\rm eq}\simeq
11\;(\Omega_0 h^2)^{-1}\mbox{\rm Mpc}$.
These dips correspond to the acoustic oscillations in the  fluid.

\newpage
\section{Comparison with previous work}

The features of the power spectrum just discussed are those expected.
We obtained the correct growth on superhorizon scales and found acoustic
oscillations in the modes which reenter the horizon in the radiation
dominated era,
as described, for example,  Bardeen {\it et al.\/} in Ref. [\ref{original}] and
in Ref. [\ref{EfstathiouReview}].
We have simply illustrated that these phenomena can be described in a
different way in the squeezed state framework.

    For a more direct comparison with the work of Grishchuk and Sidorov in
Ref. [\ref{scalarpertb}], in particular their discussion of ``desqueezing'' ,
we treat analytically a model in which matter
and radiation instantaneously decouple. We work with the action of
Eq. (\ref{eq:actionMFB}) as in [\ref{scalarpertb}]. We take
\begin{eqnarray}
a=&-\frac{1}{H\eta}\, , \;\;   \eta<\eta_2<0 \;\; \mbox{(inflation)}\\
a=&\frac{\eta}{\eta_1}\, , \;\; -\eta_2<\eta<\eta_1  \;\;\mbox{(radiation)}\\
a=&\frac{\eta^2}{(2\eta_1)^2}\, ,
\;\;\;\;\; \eta>2\eta_1 \;\;\; \mbox{(matter)}\, .
\label{eq:scalefactors}
\end{eqnarray}
The most convenient way to solve explicitly for the squeeze factor R
is to solve the equation for $v$ derived from the action
and then to use the
transformations relating the two sets of variables derived in Appendix A.
The solutions for $v$ in the three eras are
\begin{eqnarray}
v_i=&\frac{1}{\sqrt{2k}}(1-\frac{i}{k\eta})e^{-ik\eta} \\
v_r=&\sqrt{\frac{\sqrt{3}}{2k}}e^{-\frac{ik\eta}{\sqrt{3}}} \\
v_m=&\sqrt{\frac{\eta_1}{3}}((\frac{\eta}{2\eta_1})^2 +i\frac{2\eta_1}{\eta})
\label{eq:vsolutions}
\end{eqnarray}
and $\pi=v'$. The normalizations are chosen so that $v'v^*-v'^*v=-i$ in
each case. Matching $v$ and $v'$ ( and therefore $R$ and $\Phi$) continuously
at each of the two boundaries we obtain the following expressions for the
squeeze factor to leading order in $k\eta_2$:
\begin{eqnarray}
\sinh^2 R_{\vec{k}} =  {1 \over 4 (k \eta)^4 } \; , \;\; \eta<\eta_2 \\
\sinh^2 R_{\vec{k}} =  {1 \over 4 (k \eta_2)^4 }
\biggl(2-\cos\frac{2k\eta}{\sqrt 3}\biggr)\, , \;\; -\eta_2<\eta<\eta_1 \\
\sinh^2 R_{\vec{k}} =  {1 \over 4 (k \eta_2)^4 }\biggl\lbrace\;
(\alpha + \beta)\biggl(
(\frac{\eta}{2\eta_1})^4+\frac{1}{(k\eta_1)^2}(\frac{\eta}{2\eta_1})^2 \biggr)
\nonumber \\
+\gamma\biggl(\frac{\eta}{2\eta_1}
- \frac{1}{(k\eta_1)^2} \frac{\eta_1}{\eta}\biggr)
+(\alpha - \beta)\biggl(
(\frac{2\eta_1}{\eta})^2
+\frac{1}{4(k\eta_1)^2}(\frac{2\eta_1}{\eta})^4 \biggr)\;
\biggr\rbrace           \;\; \eta>2\eta_2
\label{eq:sinh2r}
\end{eqnarray}
where
\begin{eqnarray}
\alpha= \frac{1}{12}\biggl\lbrace
\biggl(5+ \frac{8(k\eta_1)^2}{3} \biggr)-\biggl(5-\frac{8(k\eta_1)^2}{3}\biggr)
\cos{2k\eta_1 \over \sqrt{3}} -\frac{4k\eta_1}{\sqrt{3}}
\sin {2k\eta_1 \over \sqrt{3}}\biggr\rbrace  \\
\beta=\frac{1}{4}\biggl\lbrace -1+\cos{2k\eta_1 \over \sqrt{3}}
+{4k\eta_1 \over \sqrt{3}} \sin {2k\eta_1 \over \sqrt{3}} \biggr\rbrace\\
\gamma= -\frac{2}{3}\biggl\lbrace
\biggl(-1+ {2(k\eta_1)^2 \over 3}\biggr) +\biggl(1+{2(k\eta_1)^2 \over
3}\biggr)
\cos{2k\eta_1 \over \sqrt{3}} -{k\eta_1 \over \sqrt{3}}
\sin {2k\eta_1 \over \sqrt{3}} \biggr\rbrace .
\label{alphabetagamma}
\end{eqnarray}

{ }From the second  expression in (\ref{eq:sinh2r}) we see that the
squeeze factor is modulated only slightly in the radiation era in agreement
with what we found earlier. For the matter era however
one can show that the coefficients $\alpha+\beta$ and $\gamma$ of the
terms which grow with $\eta$ vanish when the condition
\begin{eqnarray}
\frac{k\eta_1}{\sqrt 3}  + \arctan\frac{2k\eta_1}{\sqrt 3} = n\pi
\qquad (n\quad {\rm integer})
\label{eq:zerocondition}
\end{eqnarray}
is satisfied. This leads to a significant amount of ``desqueezing'' of these
modes as the squeeze factor for these modes is given approximately
by
\begin{eqnarray}
\sinh^2 R_{\vec{k}}=\sinh^2 R^o_{\vec{k}}
\;\biggl(\frac{2\eta_1}{\eta}\biggr)^2
{{1+\frac{1}{(k\eta)^2}} \over {1+\frac{1}{(2k\eta_1)^2}}} \, ,
\end{eqnarray}
where $R^o_{\vec{k}}=R_{\vec{k}}(2\eta_1)$ is the squeeze factor at the
decoupling. In terms of the scale factor,
\begin{eqnarray}
R_{\vec k} \simeq  R^o_{\vec k} - \frac{1}{2} ln (\frac{a}{a_{\rm dec}}) \,\;
\;\;\;{\rm for} \;\; k\eta\gg 1.
\end{eqnarray}

	The existence of this ``desqueezing'' is again a familiar phenomenon
expressed in a different set of words. When one matches the oscillating
solutions of the radiation era onto the growing and decaying solutions of
the matter era one finds that certain modes match completely onto
the decaying solution.
In fact this is the simplest way to derive the condition
(\ref{eq:zerocondition}) above. These
modes lose power and we have approximate zeros in the power spectrum. These
oscillations in the power spectrum are known as Sakharov oscillations
[\ref {SakharovZeldovich}].
In order to obtain the  position of the zeroes,
we solve Eq.~(\ref{eq:zerocondition}) and obtain
$k\eta_{\rm rec}=2k\eta_{1}=
\lbrace 6.36,\, 16.7,\, 27.4,\, 38.2,\, 49.1,\,59.9,\,70.8,\,
..\rbrace $,
which correspond to today's scales:
$\lambda=\lbrace 89,\, 34,\,20.7,\,14.8,\,11.5,\,9.5,\, 8.0,\,
..\rbrace $
\mbox{$ h^{-1}\rm Mpc$} ($h$ is the Hubble constant today in units
$100 \mbox{km/s/Mpc}$ ).
	The occurrence of these oscillations depends crucially on the matching
at the inflation-radiation transition. In order to match purely onto the
decaying solution (in the matter era) ,
one must have standing wave solutions in the radiation era
and this in turn depends on having the correct input from the inflationary
epoch. It is indeed the squeezing of all of the physical momentum out of
the superhorizon modes during inflation that produces the standing waves at
the end of inflation, which one
requires to produce this effect.

	Grishchuk and Sidorov suppose this crucial ingredient to be missing in
standard treatments of the growth of perturbations. They claim that incorrect
assumptions about the perturbations produced by inflation are often made which
lead to traveling  wave solutions in the radiation dominated era and the
resultant absence of these Sakharov oscillations in the power spectrum.
For example, Grishchuk states in [\ref{scalarperta}] that
``the unavoidable property
of squeezing manifests itself in the fact that the phases of
primordial density perturbations are fixed and correlated, {\em in contrast to
the usually made assumption} that the phases are distributed randomly and
evenly. In other words, the primordial density perturbations, similarly to the
case of gravitational waves, must form a set of standing waves with definite
phases''. [Our italics]. In fact these two points are not in conflict.
Indeed there {\em are}
``standing waves with definite phases'', but there are other phases
which are distributed randomly and evenly.  One must be careful about
which phases one is
talking about.  Each standing wave has a ``phase of oscillation''
which distinguishes among solutions which are at different points in
their period of oscillation.  This is the phase which is fixed
(relative to the time of horizon crossing) in inflationary
cosmologies.

However, inflation does {\em not} predict the location
of the nodes in the standing wave.  There is another ``spatial'' phase
which distinguishes
among standing waves which differ by a translation in space.  Since the
wavefunction assigns equal probability to solutions which differ only
by a translation, one can choose a random spatial phase.  This amounts
to making a particular random choice
of $\delta (x)$ from among the many possible ones.

We are aware of  one place in the literature where an error is made regarding
which phases are random.  In a passage
in \ref{Peebles81} (preceding the
paragraph containing Eq. (7)) Peebles argues that the temporal phase of the
standing waves may be taken to be random.
This  statement  is incorrect  and, to the extent that Grishchuk's criticisms
refer to it, we are in agreement with him.
\footnote{We are grateful to Jim Peebles for a discussion of this point.}
However, this is  an isolated error and is not of significance in either
the work of this author or others who  produce detailed predictions
based on specific models (see {\it e.g.\/} Ref. [\ref{EfstathiouReview}]).

Typically the correct standing wave solutions are used without
making reference to the
squeezed state
terminology.
That this so can most simply be seen by the fact that the usual Bunch-Davies
vacuum
matched onto the oscillating radiation era solutions gives precisely the
standing waves noted by Grishchuk and Sidorov. In
[\ref{EfstathiouReview}], for example,
the matching is described in terms of growing and
decaying modes in the radiation era, but amounts to the choice of standing
waves and indeed both the acoustic oscillations and Sakharov oscillations
which result are seen in these
simulations. The reason why so little attention is paid to these features is
that they occur only in the baryonic component of matter and are almost
completely swamped in dark matter dominated models. It is an interesting
possibility that this difference might be exploited to distinguish between
baryonic and dark matter dominated models. Attempts have in fact been made to
look for these Sakharov oscillations but the results are inconclusive
[\ref{askpeebles}].

	The other important claim of Grishchuk and Sidorov is that these
features can be said to be of a distinctly quantum mechanical origin. Speaking
of desqueezing, they state in [\ref{scalarpertb}] that ``we relate this
{\em quantum effect} to the
effect of the so-called Sakharov oscillations known in the classical theory
of matter-density perturbations''. In [\ref{scalarperta}] Grishchuk opines
that
``it is quite possible that the very specific properties of the large scale
density perturbations {\em related to their quantum mechanical origin} can
be revealed in the appropriate observations''. [Our italics].
We will attempt to clarify this question in the next section.

\newpage
\section{The Classicality of Squeezed States}

A squeezed state seems to be an especially quantum mechanical
state. It is not well localized in $p$ and $q$ and therefore
cannot be represented by a point in classical phase space. It may
instead be viewed as a {\em coherent superposition} of many localized wave
packets. It is very unlike the archetype classical state --- the coherent
state --- being very squeezed in one variable. It is this feature which
generates so much interest in these states in quantum optics and
other areas of physics and leads to their characterization as
very ``non-classical'' [\ref{quantumoptics}].
\subsection{\it Quantum coherence}
An important question is:  How can the quantum coherence of the squeezed
 state manifest
itself in physical processes?  To clarify  this question let
us make a few general remarks about quantum coherence.  A wavefunction
$\psi(x,t)$ assigns probability $\psi^{\ast}(x,t)\psi(x,t)$ to the states
$|x\rangle$.  If all one ever asked about were the probabilities
assigned to states $|x\rangle$ at time $t$, one would be working with the
equivalent of a classical probability distribution.  Quantum coherence
comes into play when one asks, for example,  about probabilities
assigned at time $t$ to states
{\em other} than $\hat x$ eigenstates.  At this point knowing
$\psi^{\ast}(x,t)\psi(x,t)$ (the probability distribution in $x$ space) is
not enough, and one needs the information provided by the complex function
$\psi(x,t)$ to generate new probabilities.  One can say that this is
because the state is a ``coherent superposition'' of $\hat x$
eigenstates.   It is possible to put the system in an {\em incoherent}
superposition of $\hat x$ eigenstates by representing its state as a
density matrix of the form $|x\rangle p(x) \langle x|$.  In this case
the probabilities $p(x)$ are all  you need to know.

In the case of the coherent superposition  $\psi
(x)$, one can avoid all
question of quantum coherence by limiting one's attention to the
probabilities assigned to the $\hat x$ eigenstates at time $t$.  However, the
nature of the time evolution can make the quantum coherence hard to
avoid.  A well known example is the double slit experiment.  If one
starts at some time $t_1$ knowing $\psi(x,t_1)$ for the electron {\em before }
it passes through the slits, then at a late time $t_2$,
$\psi^{\ast}(x,t_1)\psi(x,t_1)$
will always give
the probability assigned to whatever the initial state $|x,t_1\rangle$ evolves
into under time evolution to $t_2$.
One can avoid questions of quantum coherence at time $t_2$ by limiting
one's attention to these evolved states.
The problem is that in the double slit experiment the $|x\rangle$
states evolve into something very complicated, and one's attention (eg
measurement) is focused on other simpler states (such as states which
are eigenstates of $x$ at $t_2$).
One thus requires a knowledge not just of
$\psi^{\ast}(x,t_1)\psi(x,t_1)$, but of the full complex phase information in
$\psi(x,t_1)$.  This is when quantum coherence is important.

However, when the evolution of each basis state (with respect to
which the initial wavefunction is expanded) is simple enough, one can
realistically expect to limit one's attention to  whatever states
this initial basis evolves into. This allows one to regard the square
of the wavefunction
 as giving a classical probability distribution, and
avoid any question of quantum coherence.
A particular example of this simple
evolution is when the state is WKB classical.

\subsection{\it WKB classicality of squeezed states}
Consider the $q$ representation of the squeezed state in the static
inverted harmonic oscillator which we considered earlier
\begin{eqnarray}
\psi(q)=Ne^{-(B+iC){{q^2}\over{2\hbar}} },
\label{qrep}
\end{eqnarray}
where
\begin{eqnarray}
N =\left({B \over \hbar \pi}\right)^{1 \over 4}\, , \qquad
B ={{1}\over{\cosh 2r}}\, , \qquad
C =\tanh 2r.
\label{reln}
\end{eqnarray}

We will show that for large squeezing this wave function is {\it very}
classical in the WKB sense and becomes increasingly so with time. The
 wavefunction can be written
\begin{eqnarray}
\psi(q)=\rho(q)e^{iS(q)}.
\label{wkbform}
\end{eqnarray}
If $S(q)$ varies much more rapidly with $q$ than
$\rho(q)$ the state is a WKB state for which
\begin{eqnarray}
\hat{p}|\psi \rangle \simeq (\hbar \partial_q S(q))|\psi \rangle.
\label{ppsi}
\end{eqnarray}
To the extent that
this holds the state assigns momentum and position simultaneously
according to
\begin{eqnarray}
p(q)=\hbar \partial_qS(q).
\label{poq}
\end{eqnarray}
While $p(q)$ need not be localized, it does represent a distribution
in classical phase space which evolves classically in the WKB limit.

For the evolved <state given by Eq.~(\ref{qrep}) we have
\begin{eqnarray}
\rho(q) &=& Ne^{-B{ {q^2} \over {2\hbar}} } \\
S(q) &=& -C{{q^2}\over{2\hbar}}\, .
\end{eqnarray}

The WKB condition is met when the quantity
$\rho(\partial_qS(q)/ \partial_q\rho(q))$ is large. { }From Eq.~(\ref{qrep})
we find
\begin{eqnarray}
\left|\rho{{\partial_qS(q)} \over {\partial_q \rho(q)}}\right| = {C \over
B}=\sinh 2r.
\end{eqnarray}
Therefore as the initial state evolves and becomes more
squeezed, it also becomes more classical in the WKB sense.

Equivalently this can be seen from Eqs. ~(\ref{alpha}) --~(\ref{psquared})
since they imply
\begin{eqnarray}
| \langle \Psi(t)| \hat{q} \hat{p} |\Psi(t) \rangle |={\hbar \over
2}(1 + \sinh^2 2r)^{1 \over 2} \simeq {\hbar \over 4}e^{2
 t}.
\label{avqp}
\end{eqnarray}
This just expresses more directly the effective irrelevance of the
noncommutativity of the position and momentum operators on the state
for large squeezing.

It is precisely these properties of the inverted harmonic oscillator
which were used by Guth and Pi in [\ref{GuthPi}] to illustrate how a quantum
mechanical state can be treated in certain cases
as an ensemble of classical states.
This WKB classicality means that the squeezed state
can be approximated in its evolution as a classical phase space distribution,
as long as one only measures classical quantities.
When a particle in a spread out WKB state interacts with another
system which responds to (or ``measures'') the value of $p$ or $q$,
one can predict  the outcome using only the probability distribution
in classical phase space.  One does not need to know the complex phase
information contained in the full wavefunction, so questions of quantum
coherence do not arise.

In fact, it is well established that when such a measurement takes
place correlations are set up which cause the
quantum coherence to be lost (see for example [\ref{Albrecht}]).
{ }From that point on the
particle is in a density matrix rather than a pure state, and the
possibility of observing the effects of quantum coherence is even more
remote.

In quantum optics, where squeezed state of the electromagnetic field
can be produced, one can {\em not} think in terms of a classical
probability distribution.  This is because the electromagnetic fields
are measured by the absorption of photons by atomic systems.  These
interactions typically do {\em not} amount to a measurement of the
classical field variables, and so quantum coherence effects are observed.

The crucial  question then is: when matter  interacts with a density field
in a squeezed quantum state, does it respond to (or measure) the
classical field values or something else?  We have given this question
some thought, and find it hard to see anything other than very
classical processes in these interactions.  The matter, after all,
evolves according to the values of things such as the Newtonian
potential, which is local in the field variable.  Furthermore
as the
universe evolves the matter responds to the perturbations,
correlations will be set up which destroy the initial coherence  as
discussed above.

If one wishes to show that the initial quantum coherence of the
squeezed state is of physical importance, one must demonstrate
interactions which  measure something other than classical
quantities {\em before} the ordinary interactions destroy the quantum
coherence.  It would be very interesting if this could be done, but we
do not see how.

The particular features of the
power spectrum discussed by Grishchuk and Sidorov
are {\em not} the result of quantum coherence.
They are
features which appear in individual classical solutions (eg properties of
each trajectory  in classical phase space) and do not represent quantum
interference among different classical solutions.
The physical origin of the fluctuations (the vacuum fluctuations)
is quantum mechanical but their  known physical effects are
indistinguishable from fluctuations from a classical stochastic field.

Regarding the phases of modes which oscillate inside the horizon,
these are predicted regardless of whether there is quantum coherence.
The prediction is based on the fact that the modes in question have
spent a long time outside the horizon, where there is one growing and
one decaying solution to the equations of motion. The growing
component becomes completely dominant for the modes which are amplified
during inflation. This growing solution has a uniquely determined
oscillatory behavior when it enters the horizon, and thus the phase
of the oscillations is predicted. The original work on this subject has
correctly accounted for these predictions [\ref{original}].

The quantum squeezing is also a consequence of the presence of one
growing and one decaying solution, but that does not mean that
observing the phases of the oscillatory behavior amounts to a test of
quantum coherence. An incoherent superposition (such as would result
from the establishment of correlations with particles and photons
mentioned above) would provide the same results, as long as each mode
was dominated by the growing solution. The particular features of the
power spectrum discussed by Grishchuk and Sidorov are only
quantum mechanical in origin in the mundane sense in which all
perturbations in inflation are. The physical origin of the fluctuations
is quantum mechanical but they are in their known physical effects
indistinguishable from fluctuations from a classical stochastic field.

\section{Conclusion}

We developed the squeeze state formalism to study the growth of
cosmological perturbations. The formalism is then applied to a simple
inflationary model with baryonic matter. We discussed how the standard
features, such as acoustic oscillations and Sakharov oscillations, are
characterized in the squeeze state formalism. At late times density
perturbations are semiclassical and --- for all practical purposes --- can
be well represented by a classical probability distribution function.

Confusion can be avoided if one keeps in mind that there are three very
different phases which enter into the discussion.  {\it Firstly\/}, there is
the complex phase of the wavefunction.  To the extent that the system being
studied behaves like a classical probability distribution, this phase can
be ignored.  {\it Secondly\/}, there is the phase of oscillation of standing
waves in the density field.  These are very precisely fixed in inflationary
cosmologies and this can lead to predictable {\it Sakharov oscillations\/}
at late times.  Note that this second phase is a {\em classical} phase.
{\it Thirdly\/}, there are  classical phases for each Fourier mode which
correspond to translations in physical space.  Since the inflationary
universe assigns equal probability to density fields which differ only by a
translation, these spatial phases are random within the linear approximation.

The use of squeezed states in a cosmological setting was first
advocated and implemented by Grishchuk and Sidorov to calculate
the power spectrum of primordial
gravitational waves [\ref{GSgravwaves}].
The treatment is entirely analogous to that of
cosmological perturbations; it is possible to reduce the problem to,
again, quantizing a scalar field with a time dependent mass ($z=a$).
The power spectrum
of this scalar field exhibits oscillations on certain scales. It is
possible, as Grishchuk claims, to predict the position of the dips in
the power spectrum.
However this feature is also present in the standard Heisenberg
formalism as treated by Abbott and Harari [\ref{AbbottWise}].
The squeezed state formalism gives us an intuitive way of looking at the
generation and evolution of cosmological perturbations. However the
formalism we have developed is not restricted to cosmological
applications; the equations of motion that we have derived are quite
generic of systems with quadratic hamiltonians that can be put in the
form of Eq. (\ref{eq:Ham}).

\section{Acknowledgements}

We would like to thank Arlen Anderson, Carl Caves, David Coulson,
Leonid Grishchuk,
Jonathan Halliwell, P.J.E. Peebles,  David Salopek and
Neil Turok for useful discussions and suggestions. MJ and TP are
very grateful to Imperial College theory group for its hospitality during
their visit. The work of MJ
and TP was supported by NSF contract PHY90-21984 and the David and Lucile
Packard Foundation. The work of PF was supported by Programa Ciencia.

\vfill
\pagebreak

\appendix
\section{Relating the Heisenberg and Schr\"{o}dinger pictures}

In this appendix we show how to parametrize the
Schr\"{o}dinger picture variables in terms of the squeeze state
parameters $R_{\vec k}$,
$\Phi_{\vec k}$ and $\Theta_{\vec k}$. We then demonstrate that the
classical equations of motion for the mode functions reduce to the
evolution equations for $R_{\vec k}$, $\Phi_{\vec k}$ and
$\Theta_{\vec k}$ derived in Sec.~3 (Eq.~(\ref{eq:eomRPhiTheta})).
This shows how the Schr\"{o}dinger picture problem can be reduced to
solving the classical equations of motion with an appropriate
reparametrization.

The Heisenberg picture operators $ \hat{v}(\vec{x},\eta)$ and
$\hat{\pi}(\vec{x},\eta)$ can be written as
\begin{eqnarray}
\hat{v}(\vec{x},\eta) & =
{\cal U}^\dagger(\eta,\eta_0)\hat{v}(\vec{x},\eta_0){\cal U}(\eta,\eta_0)=
\int {{d^3k} \over {(2 \pi)^{3 / 2}}}
e^{i\vec{k}\cdot \vec{x}}(u_{\vec{k}}^{*}(\eta)a_{\vec{k}}+u_{- \vec{k}}(\eta)
a_{- \vec{k}}^{\dagger}) \nonumber \\
\hat{\pi}(\vec{x},\eta) & =
{\cal U}^\dagger(\eta,\eta_0)\hat{\pi}(\vec{x},\eta_0){\cal U}(\eta,\eta_0)=
\int {{d^3k} \over {(2 \pi)^{3 / 2}}}
e^{i\vec{k}\cdot \vec{x}}(w_{\vec{k}}^{*}(\eta)a_{\vec{k}}+
w_{- \vec{k}}(\eta)a_{- \vec{k}}^{\dagger})\, .
\label{eq:op}
\end{eqnarray}
It is now easy to show, using the Heisenberg equations of motion for
$\hat{v}$ and
$\hat{\pi}$, that the mode functions $u_{\vec{k}}(\eta)$,
$w_{\vec{k}}(\eta)$ satisfy the following Hamilton equations
\begin{eqnarray}
u_{\vec{k}}' & = & w_{\vec{k}}+{{z'} \over z}u_{\vec{k}} \nonumber \\
w_{\vec{k}}' & = & -c_s^2k^2u_{\vec{k}}-{{z'} \over z}w_{\vec{k}} \, .
\label{eq:uwHam}
\end{eqnarray}
These are the configuration and momentum variables of the classical
field theory
given by the action in Eq.~(\ref{eq:actionG}). With the initial choice of
$u_{\vec{k}}(\eta_0 )=(2k)^{-1/2}$
and $w_{\vec{k}}(\eta_0)=i (k/2)^{1/2}$, corresponding to an initial
(right moving) traveling wave, the solution to Eq.~(\ref{eq:uwHam})
are uniquely defined for all times.
At $\eta=\eta_0$ we obtain the
Schr\"{o}dinger picture operators (\ref{eq:vPiSch}).
At some later time $\eta$ we have
\begin{eqnarray}
\hat{v}_{\vec{k}}(\eta) & ={1 \over {\sqrt{2k}}}
[a_{\vec{k}}(\eta)+a^{\dagger}_{- \vec{k}}(\eta)]
\nonumber \\
\hat{\pi}_{\vec{k}}(\eta) & = -i{\sqrt{k \over 2}}
[a_{\vec{k}}(\eta)-a^{\dagger}_{- \vec{k}}(\eta)]\, ,
\end{eqnarray}
where $a_{\vec{k}}(\eta)$ and $a^{\dagger}_{- \vec{k}}(\eta)$
are the Heisenberg picture annihilation and creation operators defined by
\begin{eqnarray}
a_{\vec{k}}(\eta)  \equiv
{\cal U}^{\dagger}(\eta,\eta_0)a_{\vec{k}}
{\cal U}(\eta,\eta_0) & = & {\cal R}^{\dagger}
(\Theta_{\vec{k}})
{\cal S}^{\dagger}(R_{\vec{k}},\Phi_{\vec{k}})a_{\vec{k}}{\cal S}
(R_{\vec{k}},\Phi_{\vec{k}}){\cal R}(\Theta_{\vec{k}}) \nonumber \\
& = &
\cosh{R_{\vec{k}}}e^{-i\Theta_{\vec{k}}}a_{\vec{k}}-\sinh{R_{\vec{k}}}e^{i(\Theta_{\vec{k}} +2 \Phi_{\vec{k}})}a_{- \vec{k}}^{\dagger}
\end{eqnarray}
{ }From Eq.~(\ref{eq:op}) we then get
\begin{eqnarray}
\hat{v}_{\vec{k}}(\eta)  = &{1 \over {\sqrt{2k}}}
\Bigl[a_{\vec{k}}\left(\cosh{R_{\vec{k}}}e^{-i\Theta_{\vec{k}}}-
\sinh{R_{\vec{k}}}e^{-i(\Theta_{\vec{k}}
+2\Phi_{\vec{k}})}\right) \quad\nonumber \\ &
+a_{- \vec{k}}^{\dagger}\left(\cosh{R_{\vec{k}}}
e^{i\Theta_{\vec{k}}}-\sinh{R_{\vec{k}}}
e^{i(\Theta_{\vec{k}} +2 \Phi_{\vec{k}})}\right)\Bigr] \nonumber \\
\hat{\pi}_{\vec{k}}(\eta)  = &
-i{\sqrt{k \over 2}}\Bigl[a_{\vec{k}}\left(\cosh{R_{\vec{k}}}
e^{-i\Theta_{\vec{k}}}+\sinh{R_{\vec{k}}}
e^{-i(\Theta_{\vec{k}} +2 \Phi_{\vec{k}})}\right) \nonumber\\ &
-a^{\dagger}_{- \vec{k}}\left(\cosh{R_{\vec{k}}}e^{i\Theta_{\vec{k}}}+
\sinh{R_{\vec{k}}}e^{i(\Theta_{\vec{k}}
+2\Phi_{\vec{k}})}\right)\Bigr]\, .\quad
\label{eq:opmo}
\end{eqnarray}
Comparing Eq.~(\ref{eq:op}) with Eq.~(\ref{eq:opmo}) we can identify
the mode functions to be
\begin{eqnarray}
u_{\vec{k}}(\eta) & =
{ 1 \over {\sqrt{2k}}}\Bigl(\cosh{R_{\vec{k}}}
e^{i\Theta_{\vec{k}}}-\sinh{R_{\vec{k}}}
e^{i(\Theta_{\vec{k}}+2\Phi_{\vec{k}})}\Bigr) \nonumber \\
w_{\vec{k}}(\eta) & = i{\sqrt{k \over 2}}
\Bigl(\cosh{R_{\vec{k}}}e^{i\Theta_{\vec{k}}}+\sinh{R_{\vec{k}}}
e^{i(\Theta_{\vec{k}}+2\Phi_{\vec{k}})}\Bigr)\\
\end{eqnarray}
and these define the transformation that we seek between the
Schr\"{o}dinger picture variables and the Heisenberg picture mode functions.
It is now a matter of algebra to show that Hamilton's equations for
the mode functions (\ref{eq:uwHam}) give the equations of motion for
$R_{\vec k}$, $\Phi_{\vec k}$ and $\Theta_{\vec k}$ (\ref{eq:eomRPhiTheta}).
\vskip 0.1in

\section{ Invariance of the equations of motion for $R_{\vec k}$,
$\Phi_{\vec k}$ and $\Theta_{\vec k}$}

Here we show that for the two actions  (\ref{eq:actionMFB}) and
(\ref{eq:actionG}) differing by the total derivative term $((z'/z)v^2)'$, the
equations
of motion for $R_{\vec k}$, $\Phi_{\vec k}$ and $\Theta_{\vec k}$
have invariant form. For the action (\ref{eq:actionG})
$\lambda_{\vec{k}}$, $\Omega_{\vec{k}}$
and $\varphi_{\vec{k}}$ are defined in Eq.~(\ref{eq:OmegaLambdaPhi}).
On the other hand for the action (\ref{eq:actionMFB}) we have
\begin{eqnarray}
\lambda_{\vec{k}} & = &{k \over 2}(1-c_s^2)+{{z''} \over{2kz}} \nonumber \\
\Omega_{\vec{k}} & = & {k \over 2}(1+c_s^2)-{{z''} \over{2kz}} \nonumber \\
\varphi_{\vec{k}} & = & -{\pi \over 4}\, .
\end{eqnarray}
Even though canonically related Hamiltonians give different evolution for
$R_{\vec{k}}$, $\Phi_{\vec{k}}$
and $\Theta_{\vec{k}}$, the physically measurable quantities are
invariant.
We have not investigated how generic is the invariance of the
equations of motion (\ref{eq:eomRPhiTheta}). We leave this as an
exercise to an inquisitive reader.

%
%
\newpage
\section*{References}
\begin{enumerate}
\item
\label{original}
S. W. Hawking, {\it Phys. Lett} {\bf B\thinspace115}(1982)295.
\\
A. Starobinsky, {\it Phys. Lett.} {\bf B\thinspace117}(1982)175.
\\
A. Guth and S.-Y. Pi {\it Phys. Rev. Lett.} {\bf 49}, (1982) 1110
\\
V. Mukhanov and G. Chibisov, {\it Zh. Eksp. Teor. Fiz} {\bf 83}(1982)475.
\\
J. Bardeen, P. Steinhardt, and M. Turner, {\it Phys. Rev. D} {\bf 28} (1983)
679.
\\
{R. H. Brandenberger, {\it Nucl. Phys.} {\bf B\thinspace245}(1984)328.
\\
R. H. Brandenberger, C. Hill, {\it  Phys. Lett.} {\bf
B\thinspace179}(1986)30.
\\
Y. Nambu and M.Sasaki, {\it Progr. Theor. Phys.} {\bf 83}(1990)37}.\item
\label{Smoot}
G. F. Smoot et al., {\it Ap. J.\/}
{\bf 396}, L1\thinspace(1992).
\\
E. L. Wright et al., {\it Ap. J.\/}
{\bf 396}, L13\thinspace(1992)
\item
\label{Lubin}
T. Gaier, J. Schuster, J. Gundersen, T. Koch, M. Seiffert, P. Meinhold and
P. Lubin, {\it Ap. J.\/}
{\bf 398}, L1\thinspace(1992)
\item
\label{Gorski}
K. Gorski, R. Stompor and R. Justiewicz, CNRS preprint 1992.
\item
\label{EfstathiouFrank}
M. Davis, G. Efstathiou, C. S. Frenk and S. D. M. White, {\it Nature} {\bf
356}\thinspace
(1992)\thinspace489.
\\
G. Efstathiou, talk given at the Rutherford Meeting, December 1992.
\item
\label{scalarperta}
L. P. Grishchuk, {\it ``Quantum Mechanics of the
Primordial Cosmological Perturbations''\/}, a talk given at the
{\it Sixth Marcel Grossmann Meeting,\/} Kyoto, 1991\thinspace.
\item
\label{scalarpertb}
L. P. Grishchuk and Yu. V. Sidorov,{\it``Squeezed Quantum States in Theory
of Cosmological Perturbations''\/},in Proceedings of the Fifth Seminar
in Quantum Gravity, Moscow (1990), edited by M.A. Markov {\it et al.\/},
publ. World Scientific.

\item
\label{Bardeen}
{J. M. Bardeen,
{\it Phys. Rev.\/}~{\bf D\thinspace22} (1980)\thinspace1882\thinspace.}
\item
\label{BrandenbergerMukhanovProkopec}
For example, the squeeze state formalism has been applied to calculate the
entropy of linear cosmological perturbations in
\\
R. H. Brandenberger, V. Mukhanov
                 and T. Prokopec, Brown University
preprint, BROWN-HET-849\thinspace(1992);
\\
R. H. Brandenberger, V. Mukhanov and T. Prokopec,  {\it Phys. Rev. Lett.}
{\bf \thinspace 69}(1992)3606;
\\
T. Prokopec, Brown University preprint, BROWN-HET-861\thinspace(1992).
\item
\label{MukhanovFeldmanBrandenberger}
{V. Mukhanov, H. Feldman and R. Brandenberger,
              {\it Phys. Rep.\/} {\bf 215}, 203 (1992).}
\item
\label{GuthPi}
          A. H. Guth and S.-Y. Pi, {\it Phys. Rev. \/}~{\bf D32}
(1985) 1899
\thinspace(1982)\thinspace1110.
\item
\label{Halliwell}
J. Halliwell,{\it   Phys. Rev.\/} {\bf D\thinspace46} (1992)\thinspace1610.
\item
\label{Caves}
{C. M. Caves, {\it
Phys. Rev.\/}~{\bf D\thinspace26}\thinspace(1982)\thinspace1817\thinspace.}
\item
\label{quantumoptics}
{\it Journ. Opt. Soc. Am.\/} {\bf B\thinspace4} (10)
\\
{\it Workshop on Squeezed States and Uncertainty Relations}, proceedings of a
workshop held at the University of Maryland, March 1990, NASA Conference
Publication 3135.
\item
\label{Sidorov}
{Yu. V. Sidorov, {\it Europhys. Lett.\/}
{\bf 10\thinspace(5)}\thinspace(1989)\thinspace415.
\item
\label{GSgravwaves}
L. P. Grishchuk, {\it Sov. Phys. JETP\/}
{\bf D\thinspace40}\thinspace(1975)\thinspace409.
\\
L. P. Grishchuk and Yu. V. Sidorov, {\it Class. Quantum Grav.\/}
{\bf 6}\thinspace(1989)\thinspace L\thinspace161.
\\
L. P. Grishchuk and Yu. V. Sidorov, {\it Phys. Rev.\/}
{\bf D\thinspace42}\thinspace(1990)\thinspace3413.
\\
L. P. Grishchuk, {\it ``Relic Gravitational Waves and Limits on Inflation''\/},
Washington University preprint WUGRAV-93-1 (1993),
Bulletin Board: gr-qc@xxx.lanl.gov - 9304018.
\\
L. P. Grishchuk, {\it ``Cosmological Perturbations of Quantum-Mechanical
Origin and Anisotropy of the Microwave Background''\/},
Washington University preprint WUGRAV-92-17 (1993),
Bulletin Board: \hfil\break
gr-qc@xxx.lanl.gov - 9304001.
\item
\label{Zurek}
{W. G. Zurek and W. H. Unruh {\it Phys.Rev.} {\bf D\thinspace40} (1989)1071.}
\item
\label{CavesSchumaker}
{C. M. Caves, and B. L. Schumaker,
{\it Phys. Rev.\/}~{\bf A\thinspace31}\thinspace
(1985)\thinspace3068 and 3093\thinspace.
\\
B. L. Schumaker, {\it Phys. Rep. 135\/}\thinspace(1986)\thinspace317.}
\item
\label{Grishchuktwo}
{L. P. Grishchuk, {\it Phys. Rev. \/} {\rm D\thinspace45}
(1992)\thinspace4717.}
\\
{L. Grishchuk, H. A. Haus and K. Bergman,
{\it Phys. Rev. \/} {\bf D\thinspace46}
(1992)\thinspace1440.}
\item
\label{BirrellDavies}
{N. Birrell and P. Davies, ``Quantum Fields in Curved
              Space", (Cambridge University Press, Cambridge, 1982).}
\item
\label{EfstathiouReview}
{G.~Efstathiou, {\it Cosmological Perturbations,\/} in
               ``Physics of the Early Universe,'' {\it eds.\/} J.~A.~Peacock,
               A.~F.~Heavens and A.~T.~Davies (The Scottish University
               Summer School in Physics, Bristol, 1990).}}
\item
\label{SakharovZeldovich}
{See page 280 of Ya. B. Zel'dovich and I. D. Novikov,
{\it ``The Structure and Evolution of the Universe''},
publ. Univ. of Chicago Press.}
\item
\label{Peebles81}
P. J. E. Peebles, {\it Ap. J.\/} {\bf 248}(1981)\thinspace885.
\item
\label{askpeebles}
{T. Shanks, {\it Vistas in Astronomy} {\bf 28\/} (1985)\thinspace595.}
\item
\label{Albrecht}
{A. Albrecht, {\it  Phys. Rev.\/}
{\bf D\thinspace46} (1992)\thinspace5504,  and references therein.}
\item
\label{AbbottWise}
{L. F. Abbott and M. B. Wise, {\it Nucl. Phys.\/} {\bf B\thinspace244}
(1984)\thinspace541.
\\
L. F. Abbott and D. D. Harari, {\it Nucl. Phys.\/} {\bf B\thinspace264}
(1986)\thinspace487.}
\end{enumerate}

\newpage
\section*{Figure captions}

Fig. 1.

 Phase space trajectories for a classical upside-down harmonic
oscillator.  The presence of one growing and one decaying solution
produces a ``squeezing'' effect even at the classical level.  The
circular region shown evolves with time into the squeezed shape above it.

Fig. 2.

Evolution of scales in an inflationary universe model. $x_c$ denotes the
comoving scale and $a_R$ the end of the inflationary stage. The perturbation is
on superhorizon scales in the interval $[a(\eta_{x1},\eta_{x2}]$.

Fig. 3.

Evolution of the squeeze factor $R$ as a function of the scale
factor $log\; a$ in an inflationary universe model for two scales:
$k\eta_{\rm eq}=0.1$ and $k\eta_{\rm eq}=3$. Most of the growth occurs
on superhorizon scales (period between the marks $1x$ and $2x$).

Fig. 4.

The squeeze  phase {\it vs.\/} squeeze factor  ($\Phi - R$) diagram for two
scales:  $k\eta_{\rm eq}=1$ and $k\eta_{\rm eq}=10$. The squeeze angle freezes
out on superhorizon scales. On subhorizon scales it exhibits two types of
behavior: for scales bellow critical
($k_{\rm crit}\eta_{\rm eq}\sim 2$), the $\Phi - R$ curve exhibits oscillatory
behavior, and for scales above critical the phase remains frozen.

Fig. 5.

The growth of the power spectrum $|\delta_{\vec k}|^2$ against the scale factor
$a$ for $k\eta_{\rm eq}=0.1$ and $k\eta_{\rm eq}=3$. In both cases we observe
the same power law growth on superhorizon scales. In the subcritical case
($k\eta_{\rm eq}=0.1$) the growth continues  (with somewhat slower rate), while
in the supercritical case ($k\eta_{\rm eq}=3$) the power exhibits
oscillations  after the horizon crossing.

Fig. 6.

The snap-shot of the power spectrum ($\log |\delta_{\vec k}|^2$ -- $\log k$
plot) for two times: $a=0.1 a_{\rm eq}$ and  $a=0.6 a_{\rm eq}$. The
spectrum is scale invariant on superhorizon scales:
$|\delta_{\vec k}|^2\sim k$ and after the turning point at
$k\eta_{\rm eq}\simeq 4$ it shows oscillatory behavior and decays as
$|\delta_{\vec k}|^2\sim k^{-1}$.

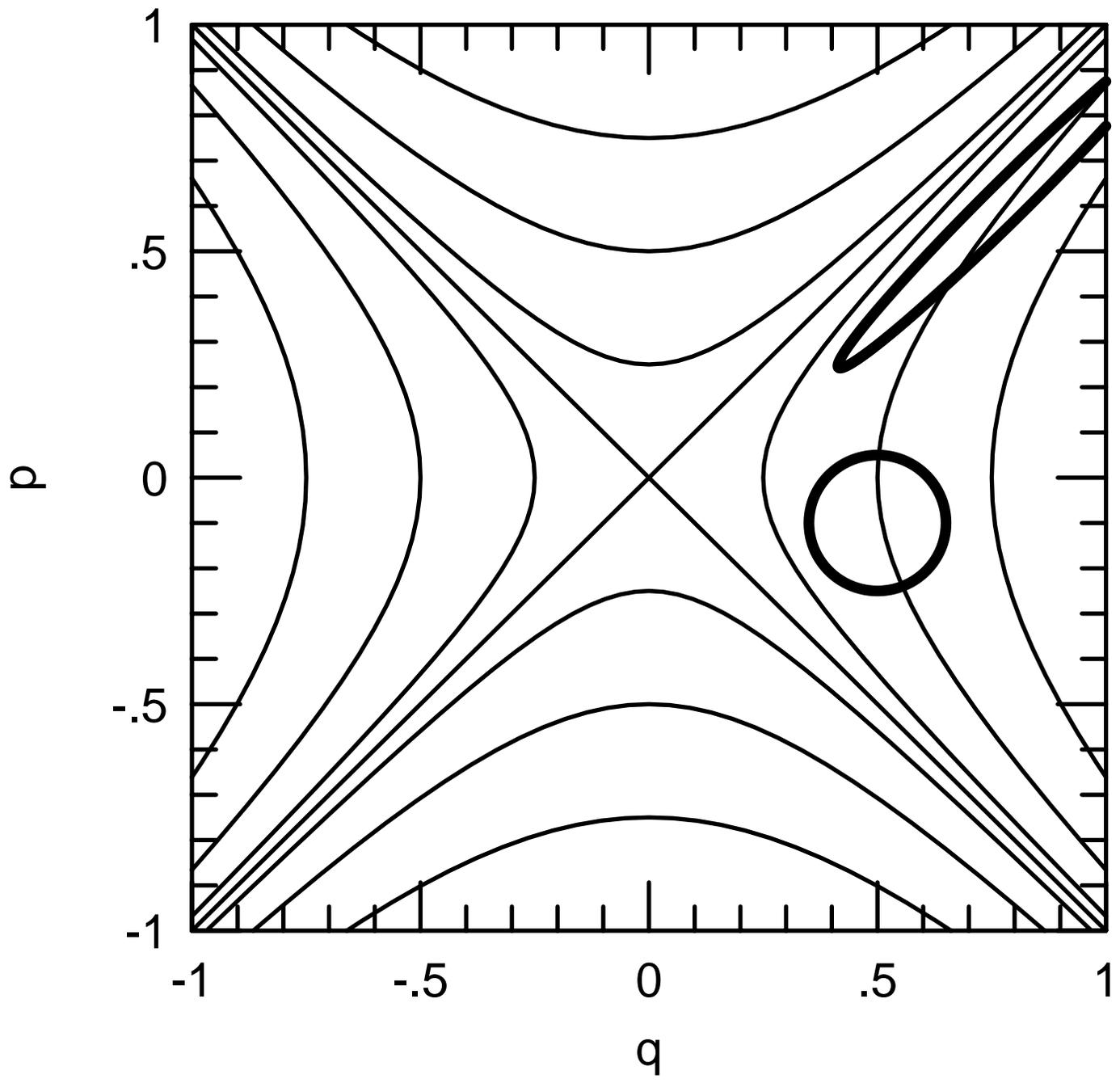
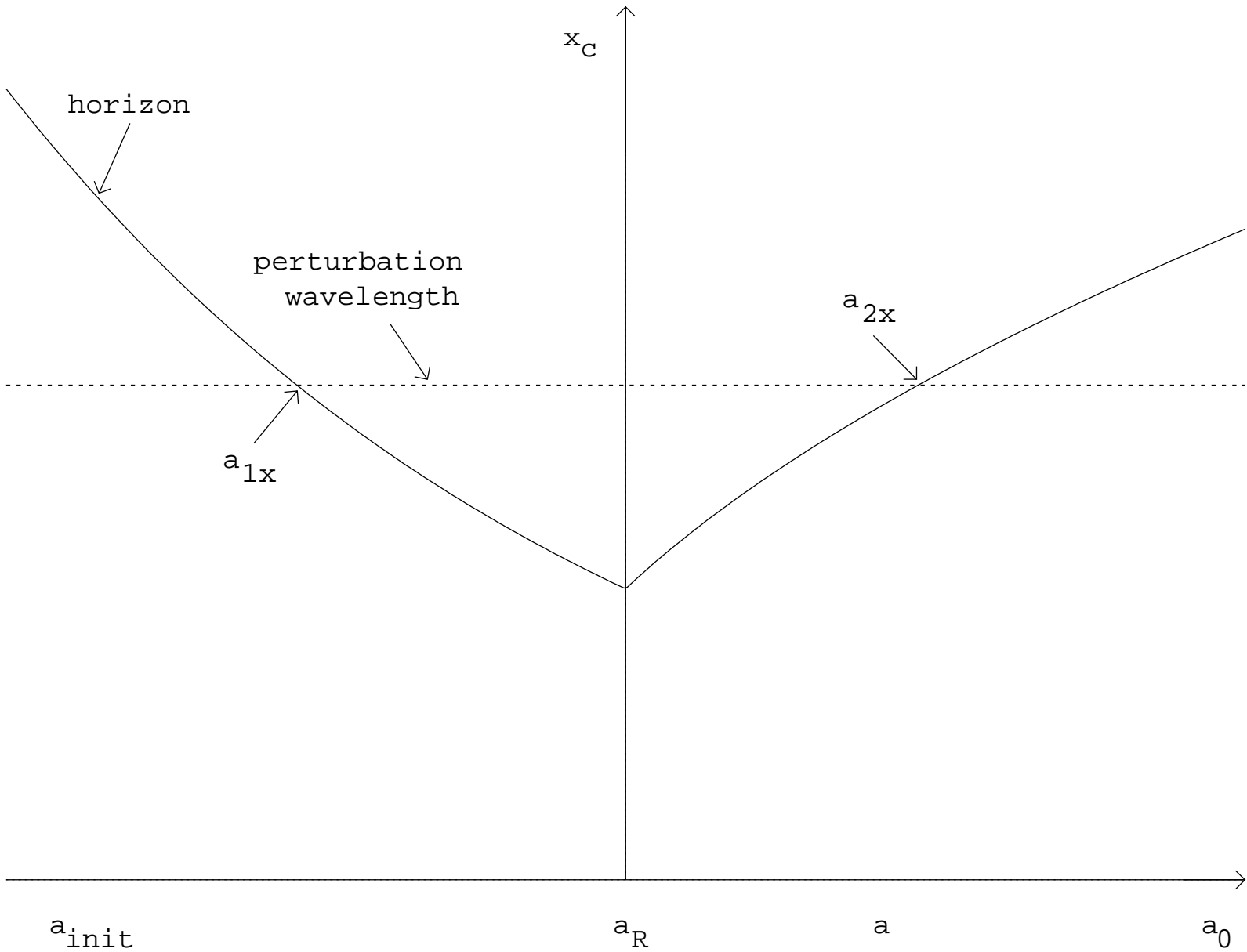
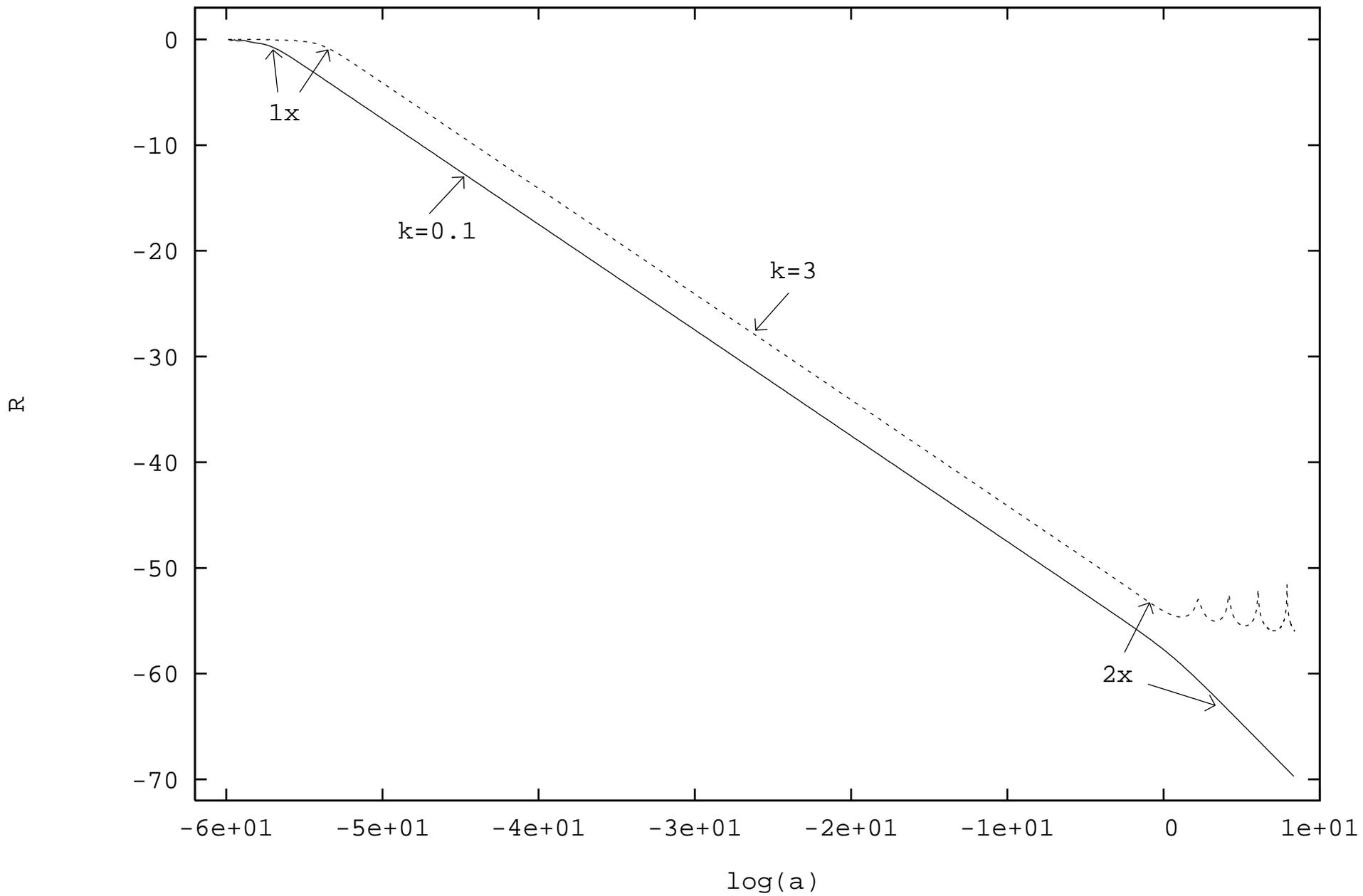
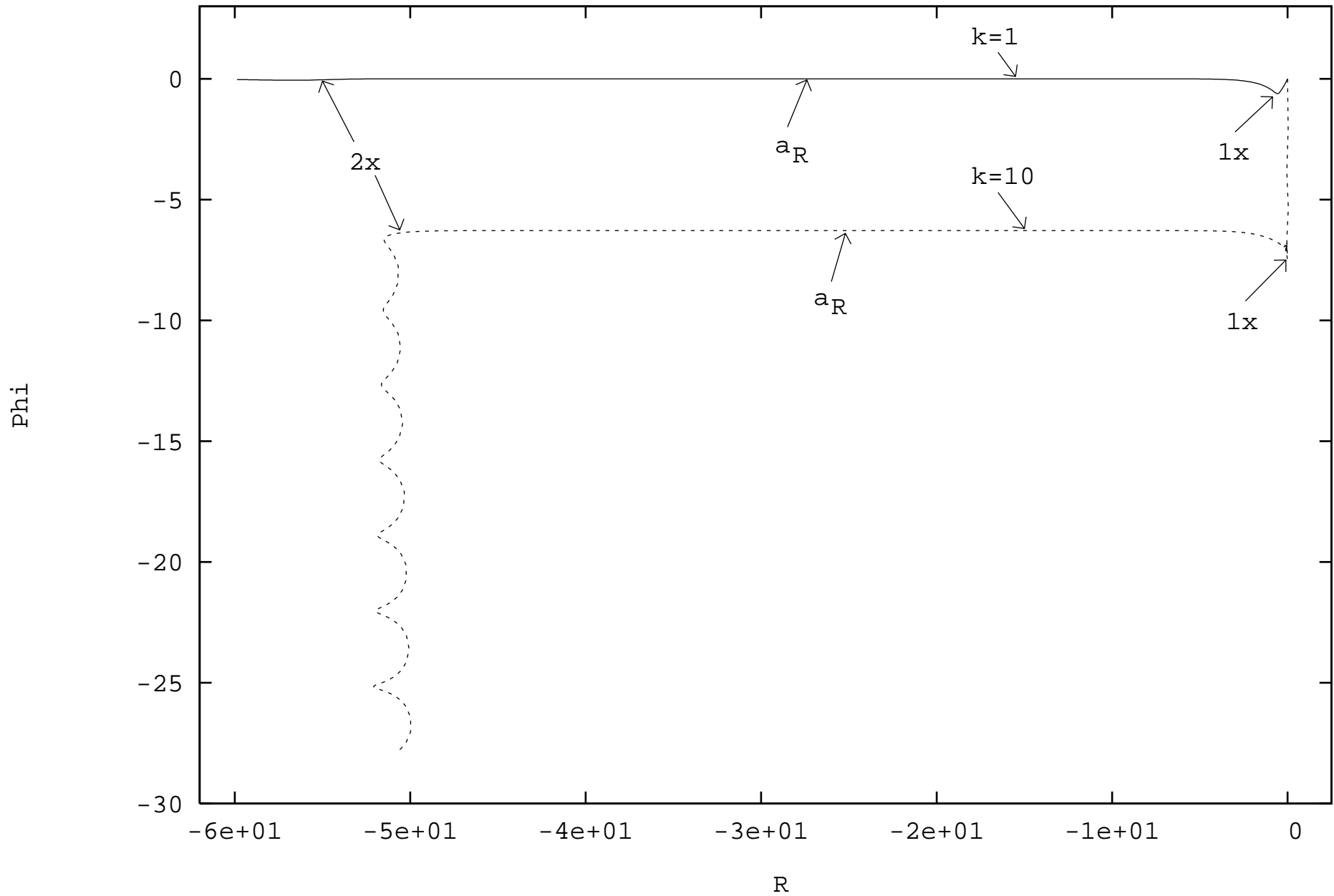
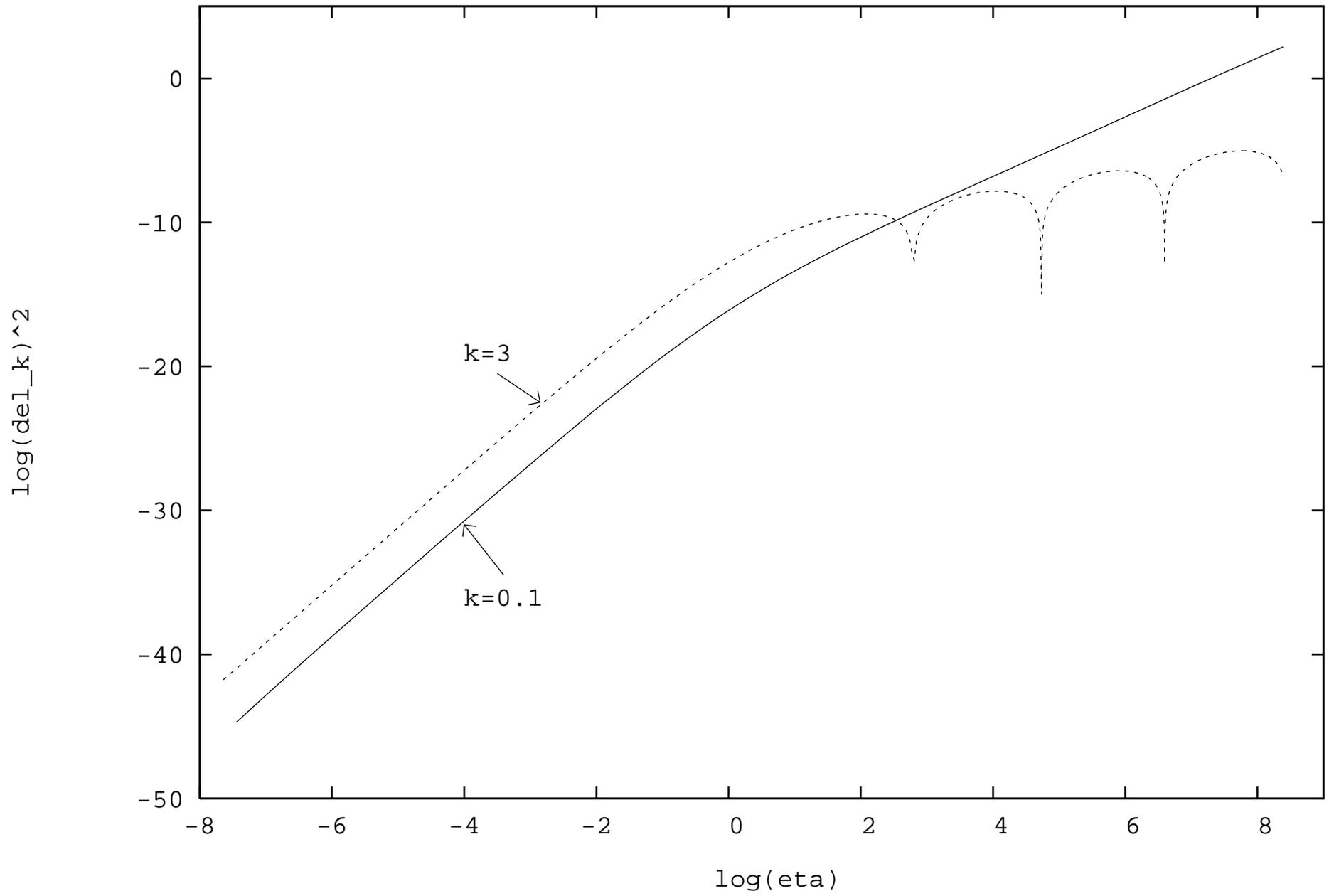
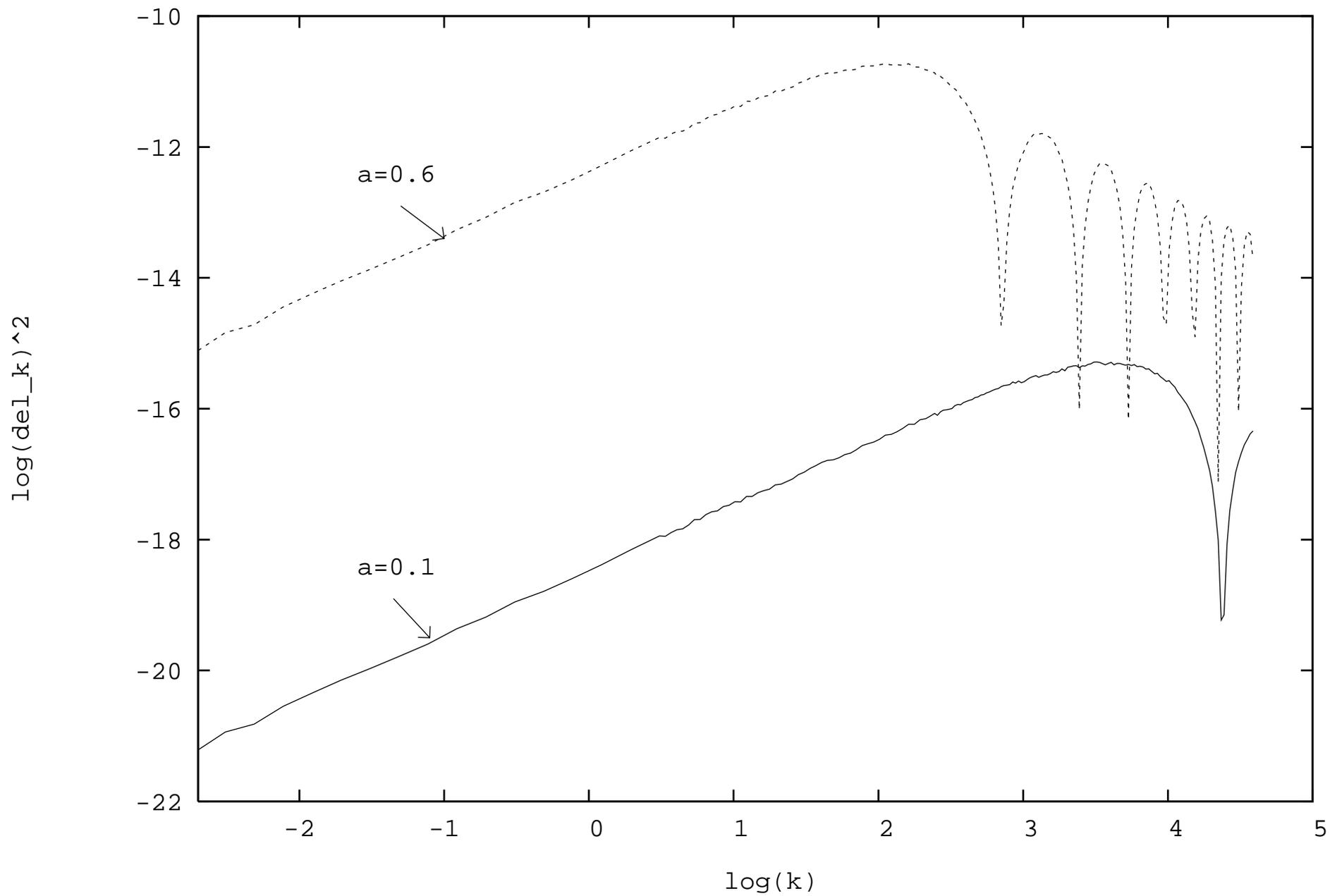
\end{document}